\documentclass[subscriptcorrection,upint,varvw,barcolor=Goldenrod3,mathalfa=cal=euler,balance,hyphenate,french,pdf-a,nolists,table,xcdraw]{asmejour} %

\usepackage[switch]{lineno} 
\usepackage{lipsum} 

\usepackage[noend]{algpseudocode}
\usepackage{algorithm}
\usepackage{algorithmicx}
\usepackage{etoolbox}

\newcommand{\algstrut}[1][\algruledefaultfactor]{\vrule width 0pt
depth .25\baselineskip height #1\baselineskip\relax}

\usepackage{enumitem}

\hypersetup{%
	pdfauthor={Saeid Bayat},                       		   	
	pdftitle={Multi-split configuration design for fluid-based thermal management systems},                  	
	pdfkeywords={Design Synthesis, Graph, Optimization, Optimal Flow Control, Thermal Management System},
	pdfsubject = {Describes the asmejour LaTeX template},			
	pdflicenseurl={https://ctan.org/pkg/asmejour},
}

\AtBeginEnvironment{algorithmic}{\lineskip0pt}

\newcommand*\Stateh{\State \algstrut[1]}
\usepackage{enumitem}
\usepackage{graphicx}
\setlist[enumerate,1]{label=\arabic*}
\setlist[enumerate,2]{label=\theenumi.\arabic*}
\setlist[enumerate,3]{label=\theenumii.\arabic*}

\begin{document}

\SetAuthorBlock{Saeid Bayat\CorrespondingAuthor}{%
Department of Industrial and Enterprise Systems Engineering,\\
University of Illinois at Urbana-Champaign,\\
Urbana, IL, USA\\
email: bayat2@illinois.edu
}

\SetAuthorBlock{Nastaran Shahmansouri\CorrespondingAuthor}{
Autodesk Research,\\
661 University Ave,\\
Toronto, Canada\\
email: nastaran.shahmansouri@autodesk.com} 

\SetAuthorBlock{Satya RT Peddada}{
Department of Industrial and Enterprise Systems Engineering,\\
University of Illinois at Urbana-Champaign,\\
Urbana, IL, USA\\
email: speddad2@illinois.edu} 

\SetAuthorBlock{Alexander Tessier}{
Autodesk Research,\\
661 University Ave,
Toronto, Canada\\
email: alex.tessier@autodesk.com} 

\SetAuthorBlock{Adrian Butscher}{
Autodesk Research,\\
661 University Ave,\\
Toronto, Canada\\
email: adrian.butscher@autodesk.com} 

\SetAuthorBlock{James T Allison}{
Department of Industrial and Enterprise Systems Engineering,\\
University of Illinois at Urbana-Champaign,\\
Urbana, IL, USA\\
email: jtalliso@illinois.edu} 
   
\title{Multi-split configuration design for fluid-based thermal management systems}


\begin{abstract}
High power density systems require efficient cooling to maintain their thermal performance. Despite this, as systems get larger and more complex, human practice and insight may not suffice to determine the desired thermal management system designs. To this end, a framework for automatic architecture exploration is presented in this article for a class of single-phase, multi-split cooling systems. For this class of systems, heat generation devices are clustered based on their spatial information, and flow-split are added only when required and at the location of heat devices. To generate different architectures, candidate architectures are represented as graphs. From these graphs, dynamic physics models are created automatically using a graph-based thermal modeling framework. Then, an optimal fluid flow distribution problem is solved by addressing temperature constraints in the presence of exogenous heat loads to achieve optimal performance. The focus in this work is on the design of general multi-split heat management systems. The architectures discussed here can be used for various applications in the domain of configuration design. The multi-split algorithm can produce configurations where splitting can occur at any of the vertices. The results presented include 3 categories of cases and are discussed in detail. 
\end{abstract}

\date{Version \versionno, \today}

\maketitle 


\section{Introduction}
\label{Sec: Introduction}
Electrification of many systems in various domains has increased over the last few decades \cite{Rootzn2020ElectrifyEC, ElRefaie2016GrowingRO, Nadel2019ElectrificationIT}. These electronic devices are experiencing intensifying miniaturization \cite{mathew2022review,bayat2021observation, smoyer2019brief, wong2020density,bayat2019observer, ye2019last}, resulting in higher power densities. Currently, microprocessors generate heat fluxes of over 102 W/cm$^2$. Hot spots on microelectronic chips can also generate heat fluxes of 1 kW/cm$^2$ or more, which result in excessive temperatures in local regions \cite{mathew2022review}. When devices are operated at excessive temperatures, their performance and reliability are negatively affected, resulting in their malfunction in the end \cite{He2020ThermalMA, Almubarak2017TheEO}. Consequently, the optimal design of thermal management systems is essential for expediting design processes and achieving ambitious performance goals \cite{Mathew2021ARO, Feng2019MultifunctionalTM}.

Throughout the literature, the design of heat management systems has largely focused on improving individual components \cite{peddada2019optimal,panjeshahi2009optimum}. Yet, optimizing individual components can result in a sub-optimal overall design. Additionally, the design of a whole heat management system sometimes is only relevant to a particular application \cite{muller2016energy}. Peddada et al.~\cite{peddada2019optimal} took an important step towards designing a class of heat management systems that are not limited to a specific application. However, the systems considered in that work are restricted to single-split system architectures, i.e., configurations with a single source where thermal-devices of the system are located in branches that only start from the source. There are, nevertheless, many instances in which multi-split systems are required \cite{Ling2016ExperimentalIO}. In a multi-split architecture, branching can occur from the main source or any thermal-device of the system. In this paper, we introduce a strategy for the generation and the optimal design of general multi-split heat management systems using graphs. Here, spatial considerations are also incorporated into the design problem; the spatial data of the heat-devices of the system are used for clustering and defining junctions where branching starts. After forming the clusters, for each cluster, we find the Euclidean distance of cluster's heat-devices from the cluster's centroid; the heat-device closest to the centroid is defined as the cluster's junction, refer to section ~\ref{Sec: Generation Multi-split Spatial Graphs} for more details. The architectures discussed in the paper can be used across a wide range of applications. Additionally, the design strategy presented here applies to both single-split and multi-split architectures, supporting quantification of trade-offs between multi-split system performance improvement and cost and complexity increases.  

Designs of these systems should meet transient response conditions   \cite{Bennett2016TheIO} as many electronic devices work in applications with time-varying workloads \cite{mathew2022review}. Accurate modeling of the design problem as an optimization problem, therefore, requires treatment of system dynamics, including bounds and constraints on dynamic behavior, as well as optimal control for active systems. Continuous optimization can be applied to the combined design of physical and control system design, a well-established problem in Control Co-Design (CCD) \cite{allison2014, allison2014-guo, garcia2019-CCD, SaeidBayat-Vehicle}, for a given system configuration. Distinct configurations can have fundamentally different dynamics, design variables, and constraints, so each configuration must be treated independently. Practical solution of such problems requires automatic generation and solution of the optimization problem for each configuration. Here, for every generated unique and feasible multi-split system configuration a continuous optimization problem is formulated and solved \cite{herber2017-enum, herber2019-combined, peddada2019optimal}. 

Thermal management system configuration design considered in this paper belongs to an especially challenging class of optimization problems where discrete decisions change the set of continuous decisions to be made. This necessitates a nested discrete/continuous approach. Furthermore, the very general nature of these problems prevent the use of established integer programming methods that can be used for efficient solution of other problem classes with spatial properties. At least, there are three possible approaches for solving general physical systems configuration design problems. The first approach, as described above and used in the studies presented here, entails enumerating all unique and feasible configurations for a given design space, solving the continuous optimization problem for each one, and then producing a ranked set of design candidates. This is the only approach that can produce a result that is a confirmed optimum. A second approach employs a population-based optimization algorithm, such as a Genetic Algorithm (GA), or other gradient-free search strategy to navigate the configuration space while still solving the continuous optimization problem for each configuration. A third approach is to utilize machine learning or artificial intelligence techniques to restructure the system configuration design space \cite{allison2022artificial, guo2019circuit} such that it is more tractable, while again nesting continuous optimization within this search. Recent machine learning strategies have proven to be only marginally better than random search \cite{guo2019circuit}. General solution approaches for configuration problems too large for enumeration and without special problem structure remains elusive, and is truly an engineering design grand challenge. 

The main contributions of the work presented in this article are:
\begin{enumerate}
    \item Introduction of a new automated modeling strategy for multi-split thermal management systems. This strategy has applications for configuration generation across a wide range of systems. 
    \item Employing spatial information to cluster data and to define junctions.
    \item Comparison of multi-split versus single-split configurations, providing insight into the trade-offs between system performance and system complexity
    \item In-depth study of multi-split system optimal results and analysis of the system signals such as flow rates and flow temperatures at heat-device locations that led to this optimal solution
\end{enumerate}

This article continues as follows, in Section \ref{Sec: System Description and Modeling}, we discuss the thermal management system architectures studied in this work. The dynamic graph based models of the thermal architectures are presented in Section \ref{Sec: Dynamic Graph-Based Modeling}. Section \ref{Sec: Generating Multi-split Spatial Graphs} describes the graph-based representation of multi-split architectures. The formulation of the variable time horizon dynamic optimization problem is explained in Section \ref{Sec: Optimal Flow Control Problem}. Section \ref{Studies} presents case studies using different architectures and heat loads. Section \ref{Conclusion} concludes with a summary of the design methodology, guidelines for thermal management system design, and potential future research topics.

\section{System Description and Modeling}
\label{Sec: System Description and Modeling}
\begin{figure}[ht!]
    \centering
    \includegraphics[width=1.0\linewidth]{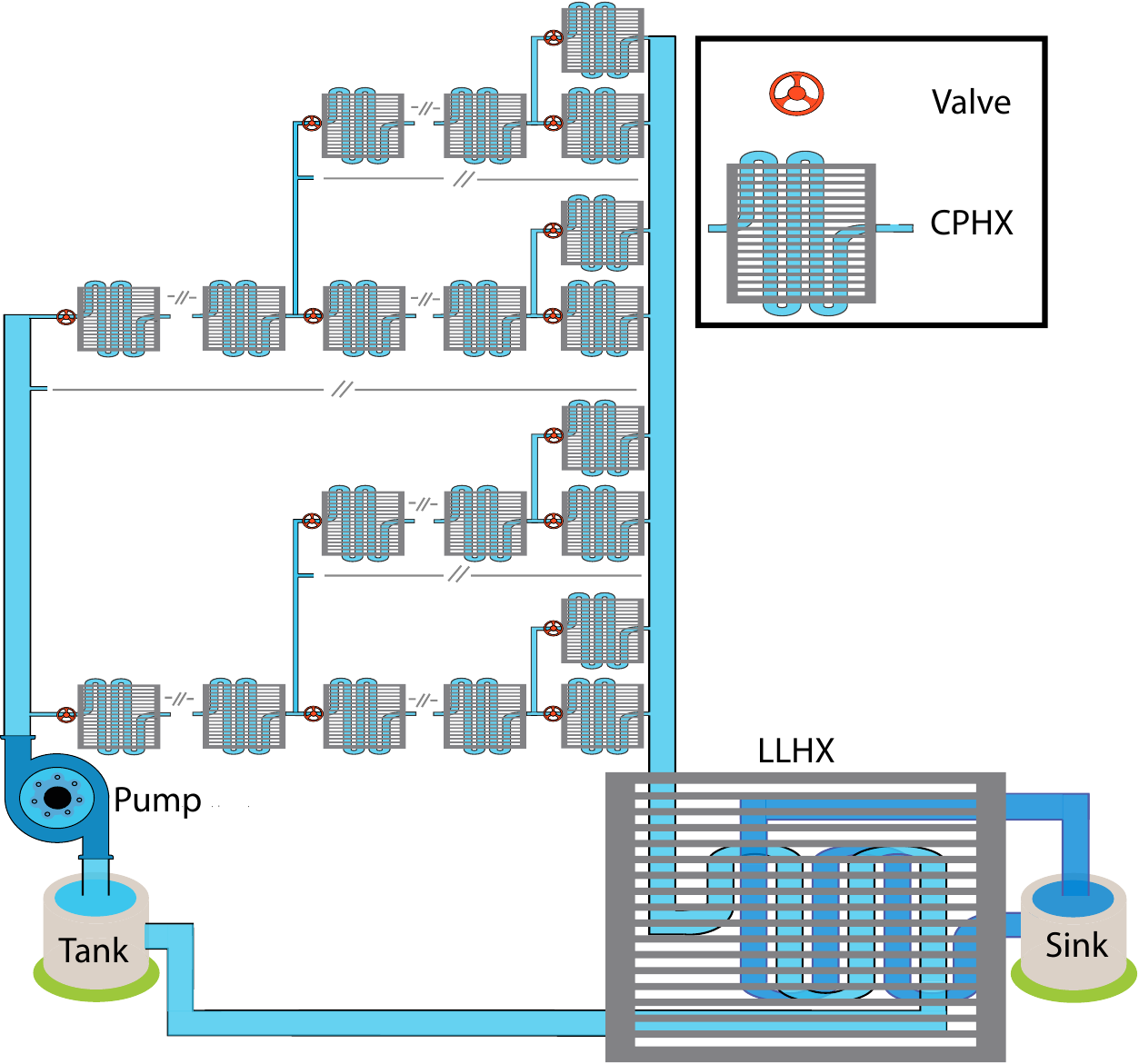}
    \caption{Class of problems considered in this paper. The systems include a tank, a pump, valve(s), CPHXs in parallel and series, a LLHX, and a sink.}
    \label{Fig:General_system}
\end{figure}

Figure~\ref{Fig:General_system} illustrates the class of problems considered in this article. The purpose of this system is to manage the temperature of various heat generating devices mounted on Cold Plate Heat Exchangers (CPHXs) through which a coolant flows. The coolant is stored in a tank and is transferred to each branch by a pump. Each branch has several valves and can divide the flow it receives into its sub-branches. The coolant that passes through heat exchangers absorbs heat and transfers it to the thermal sink through a Liquid-to-Liquid Heat Exchanger (LLHX).

The class of architectures considered in this work produces multi-split configurations as opposed to the single-split configurations generated in the study by Peddada et al.~\cite{peddada2019optimal}. A single-split assumption limits the configuration search space; expanding the search space to include multi-split configurations may enhance system performance. The multi-split enumeration algorithm created in this work produces configurations where splits may be made at the pump (source) or at any of the CPHX locations. Sections \ref{Sec: Dynamic Graph-Based Modeling} and \ref{Sec: Generating Multi-split Spatial Graphs} describe the graph-based physical modeling of the thermal systems and the representation and generation of multi-split architectures. Here the heat load applied to each CPHX, as well as the inlet temperature and mass flow rate of the thermal sink, have been assumed to be known exogenous inputs. In addition, heat generating devices are assumed to have the same temperature as the wall of the CPHX on which they are mounted. It is assumed that heat loss through pipes is zero.

An optimal control problem is defined for each fluid-based thermal management systems configuration generated using the multi-split algorithm. The control problem seeks the optimal flow rate trajectory for each pipe that maximizes system performance, while satisfying component temperature constraints. The flow rates are controlled by valves. Dynamic system models incorporate the thermal physics of advection, convection, and bi-directional advection. A unique model is generated for each configuration design. We assume that system performance is quantified by thermal endurance, i.e.,  the goal is to maximize the time that the device is on while ensuring that all temperature bounds are met. This is consistent with the approach in the study conducted by Peddada et al. ~\cite{peddada2019optimal}, supporting direct comparison with earlier single-split studies using graph-based configuration representations. 

\begin{figure}[ht!]
    \centering
    \includegraphics[width=1.0\linewidth]{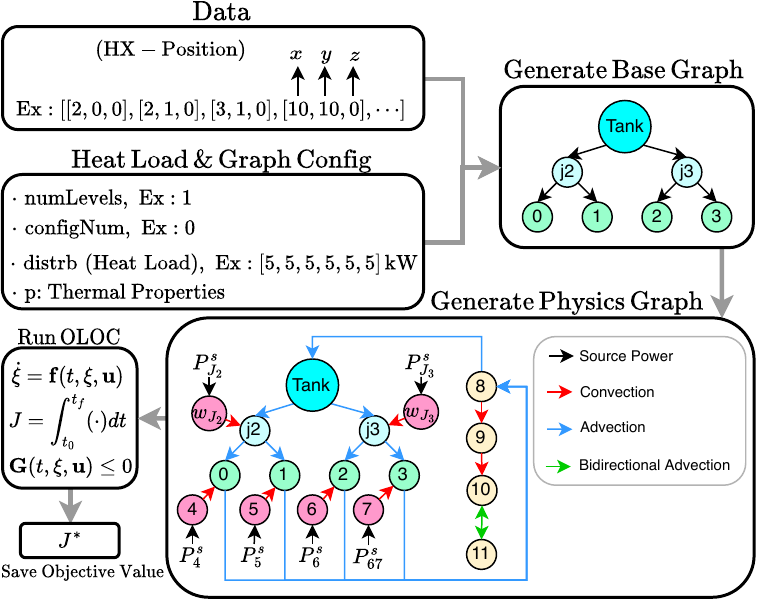}
    \caption{The workflow diagram that illustrates the steps involved in the code execution. Starting with the given \textit{Data}, \textit{Head Load}, and \textit{Config Number}, the base graph and physics graph are generated. The OLOC problem is then defined and solved to obtain the objective function value. The objective value represents thermal endurance, with the incorporation of a penalty function for control signal smoothness and convergence. Refer to Sec.~\ref{Sec: Optimal Flow Control Problem} for a comprehensive discussion}
    \label{Fig:Workflow}
\end{figure}

The code workflow is illustrated in Fig.~\ref{Fig:Workflow}, providing an overview of the code structure. In the subsequent sections, each part will be discussed in detail. The variable \textit{Data} represents the positions of CPHX in [$x,y,z$] coordinates. The parameter \textit{numLevels} determines the depth of the graph considered for junction creation. For example, when set to 1, junctions can only be added to branches connected to the tank. Multiple graphs are generated based on the provided \textit{Data} and \textit{numLevels}, resulting in different configurations, such as parallel, series, or a combination thereof. Among a set of $N$ graphs, the variable \textit{configNum} specifies a particular graph. The variable \textit{distrb} denotes the heat load at each node. Using the given \textit{Data}, \textit{Heat Load}, and \textit{Graph Config}, the \textit{Base Graph} is generated, representing the connections between the tank, junctions (denoted as $j_2$ and $j_3$ in Fig.\ref{Fig:Workflow}), and other CPHX nodes. Subsequently, a \textit{Physics Graph} is created by adding CPHX wall nodes, sink nodes, and source power (heat load) to the \textit{Base Graph}, along with the underlying physics between these nodes, which include convection, advection, and bidirectional advection. Based on the \textit{Physics Graph}, an Optimal Open Loop Control (OLOC) is defined, consisting of three components: system dynamics ($\mathbf{f}$), objective function ($J$), and path constraints ($G$). Here, $t$ represents time, $\mathbf{\xi}$ denotes the state, and $\mathbf{u}$ represents the control signal. These components will be discussed in detail in subsequent sections.

\section{Dynamic Graph-Based Modeling}
\label{Sec: Dynamic Graph-Based Modeling}
\begin{figure}[ht!]
    \centering
    \includegraphics[width=1.0\linewidth]{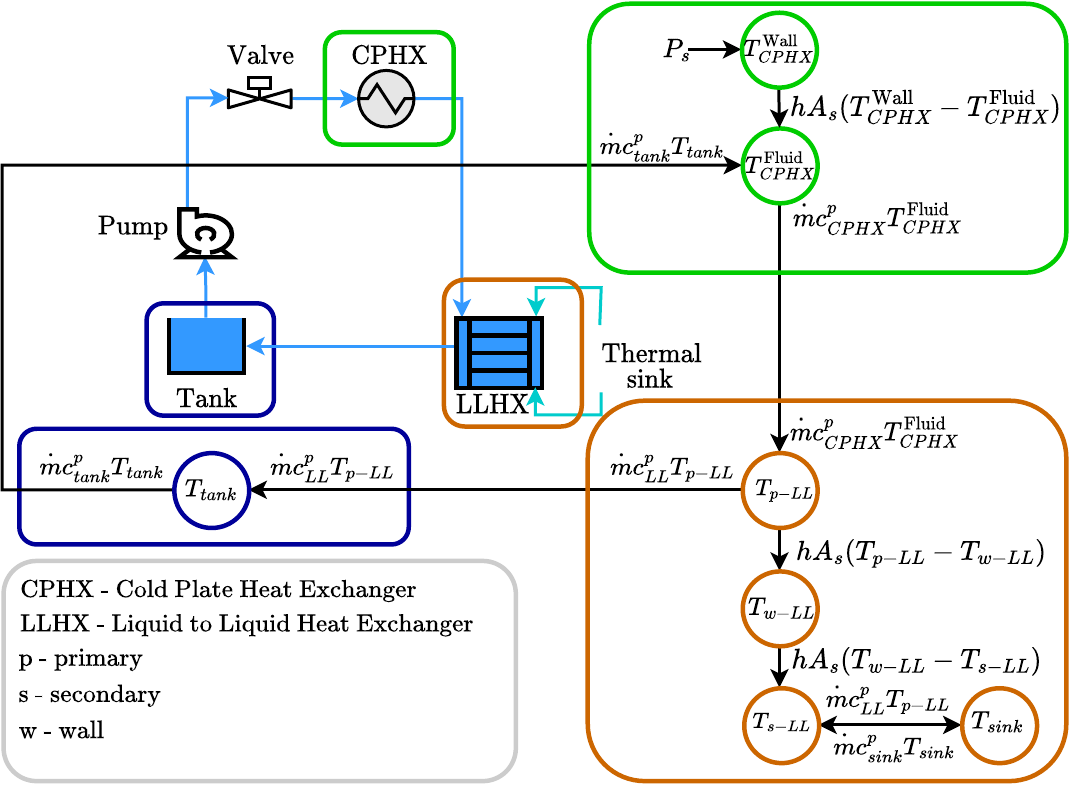}
    \caption{Notional example to illustrate the elements of thermal physics included in this system model. Here, a node represents a temperature and an edge represents a power flow.}
    \label{Fig:Graph_Modelling}
\end{figure}

This paper uses the dynamic graph-based modeling framework discussed in Peddada et al.~\cite{peddada2019optimal}. Nodes represent components or fluids, and a temperature is associated with each node. Edges represent power-flows between nodes. For brevity, model details are omitted here, but the dynamics governing the main components are presented in Fig.~\ref{Fig:Graph_Modelling}. The model includes advection, convection, and bidirectional advection. Convection happens between CPHX wall nodes and fluid nodes, advection occurs between fluid nodes, and bidirectional advection occurs between the LLHX and sink. In Fig.~\ref{Fig:Graph_Modelling}, $c^p$ is the specific heat capacitance, $\rho$ is the density of the fluid, $A_s$ is the convective surface area, $h$ is the heat transfer coefficient, and $\dot{m}$ is the mass flow rate of the fluid.

\subsection{Heat-transfer Model of Graph Nodes}
\label{Sec: Heat-transfer Model of Graph Nodes}

The power-flow type in the tank is advection, which occurs between 1) the tank and the fluid node on the CPHX, and 2) the tank and the primary side of the LLHX. Advection and convection are present in the CPHX. Advection occurs between 1) the CPHX fluid node and tank and 2) the CPHX fluid node and primary side of the LLHX. Convection occurs between the CPHX fluid and the wall node. The LLHX involves advection, convection, and bi-directional advection. Advection occurs between 1) the primary side of the LLHX and tank, 2) the primary side of the LLHX and the CPHX fluid node. Bi-directional advection takes place between the secondary side of the LLHX and the sink. Convection occurs between 1) the LLHX wall node and the primary side of LLHX and 2) the LLHX wall node and the secondary side of LLHX. These are building blocks of any larger system. Using these components, a model of the dynamics of any complex system of the form illustrated in Figure~\ref{Fig:General_system} can be generated.

\subsection{Graph-Based Model for Multi-split Architectures}
\label{Sec: Graph-Based Model for the Class of Architectures}
\begin{figure}[ht!]
    \centering
    \includegraphics[width=1.0\linewidth]{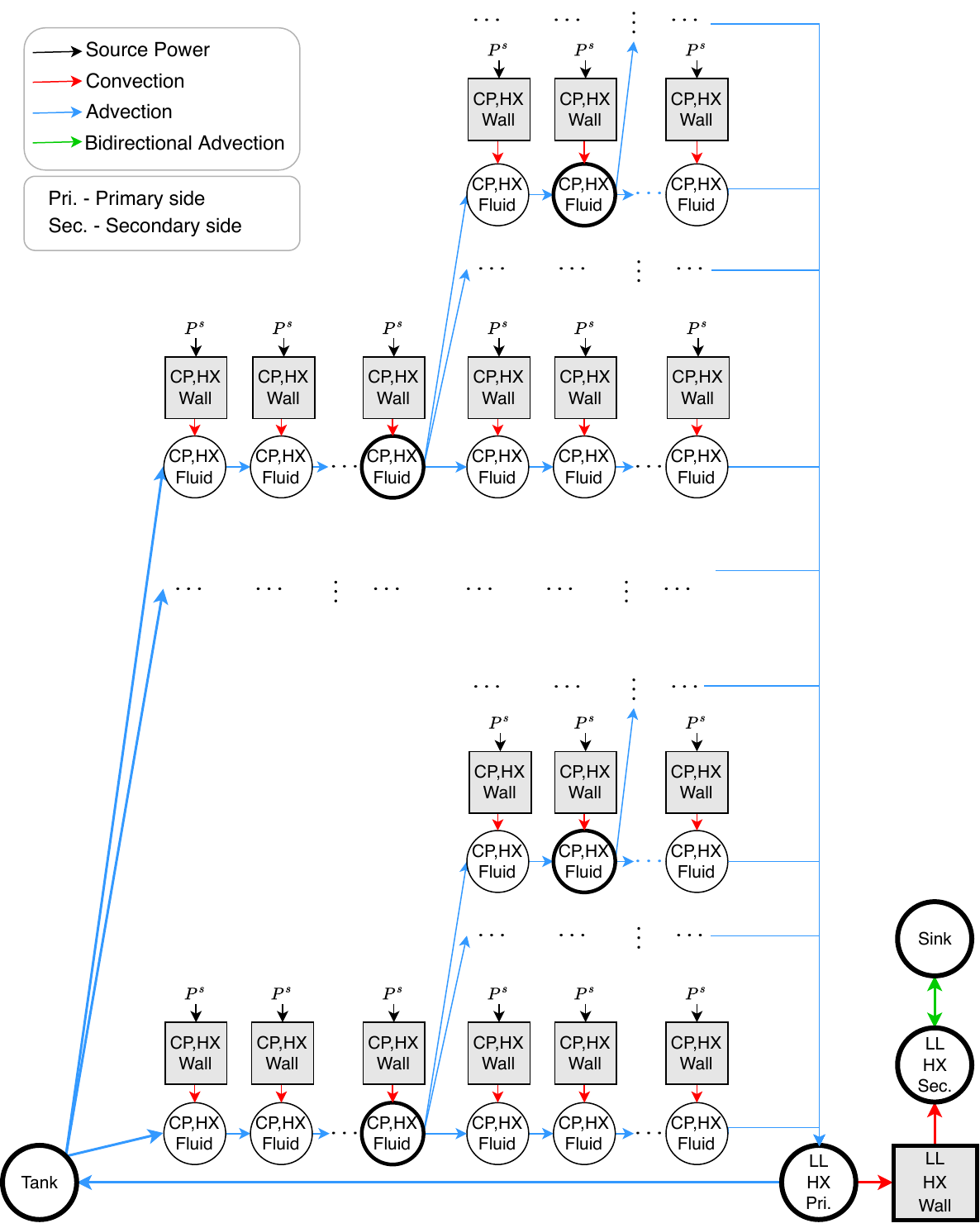}
    \caption{Graph-based model for multi-split architectures studied in this article.}
    \label{Fig:General_Graph}
\end{figure}

\begin{figure}[ht!]
    \centering
    \includegraphics[width=1.0\linewidth]{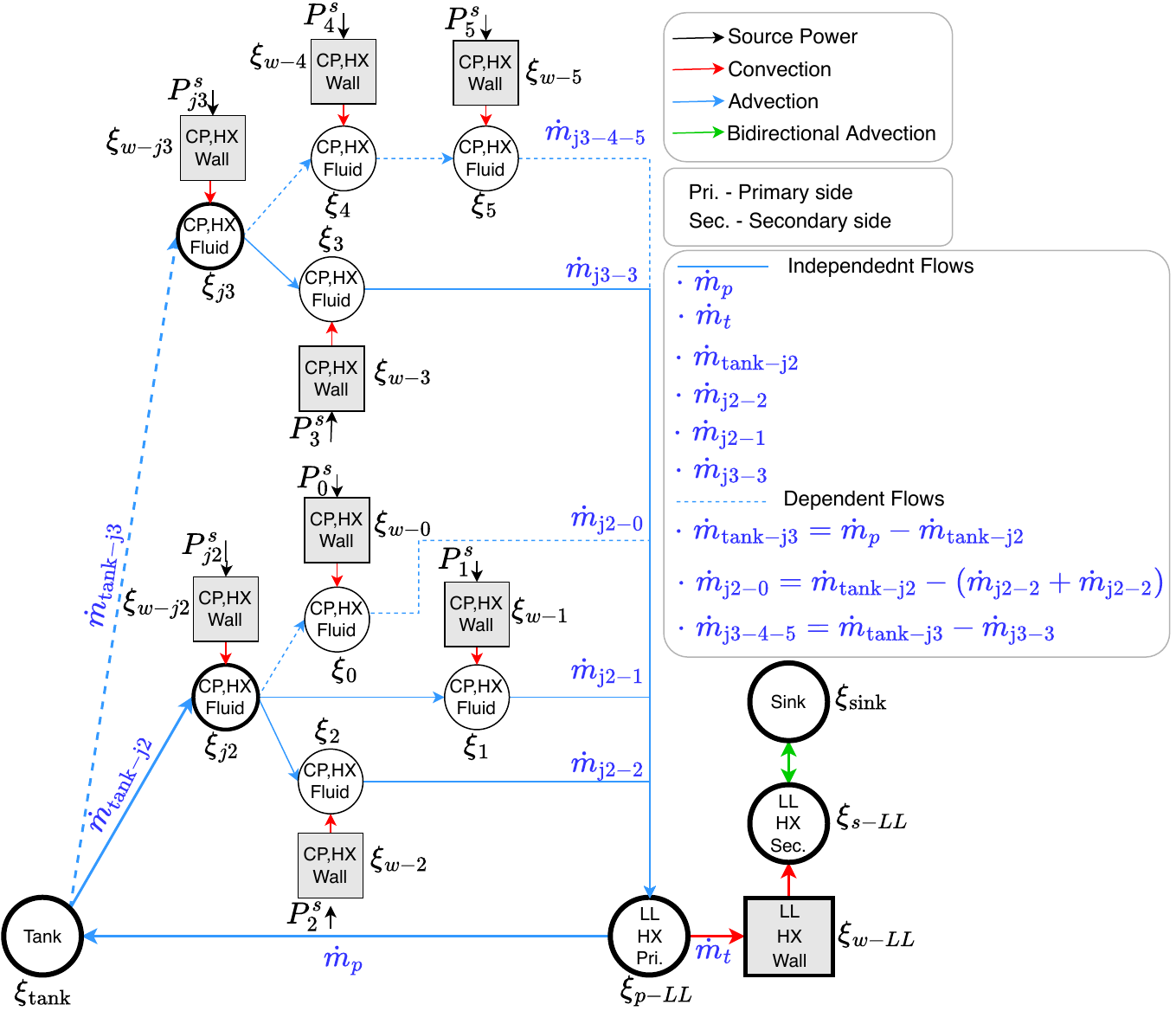}
    \caption{Graph illustrating the variables for a simple example multi-split configuration. Types of power flow are illustrated, including the distinction between independent and dependent flows.}
    \label{Fig:Graph Signal}
\end{figure}
Figure~\ref{Fig:General_Graph} shows a graph that represent the general class of multi-split architectures studied in this paper. Here, $P^s$ represents the heat load. A simple example problem of this class is shown in Fig.~\ref{Fig:Graph Signal}. Each of the two junctions has 3 associated nodes. The first junction ($j_2$) distributes the flow to 3 different branches, and the second junction ($j_3$) distributes the flow to two branches, where 2 nodes in a branch are in series, and the second branch has only 1 node. In this figure, \textit{states} are denoted using the symbol $\bm{\xi}$; here the states all correspond to node temperatures. For example, $\xi_1$ is the temperature of node 1 and $\xi_{w-1}$ is the wall temperature of the CPHX connected to node 1. The system involves both independent and dependant flows, depicted in Fig.~\ref{Fig:Graph Signal}. Independent flows can be controlled, whereas dependent flows can be calculated based on the independent flows (see Fig.~\ref{Fig:Graph Signal} for these equations). In the figure, the edges that carry dependent power flows are shown with dashed-lines.

\subsection{State Equations for the Graph-Based Model}
\label{Sec: State Equations for the Graph-Based Model}
The state equations for the thermal management systems are obtained by using a methodology similar to the one introduced in references ~\cite{peddada2019optimal, koeln2016experimental}. The methodology described there has been extended to accommodate multi-split configurations. Equation~\eqref{Eq: Dynamics} shows the state equation for this system. Here $\mathbf{T}(t)$ is the vector of node temperatures (states), $T^t$ represents the sink temperature, which is considered to be a constant value and is known, $\dot{m}_p(t)$ is the pump flow rate, $\dot{m}_t(t)$ is the sink flow rate, $\dot{\mathbf{m}}_f(t)$ is composed of the sub-branch flow rates, and $\mathbf{P}^s(t)$ is the vector of component heat loads. Here, $\mathbf{C}$ is the diagonal matrix of heat capacitance, and $\mathbf{D}$ represents the connection of external sources to the system. Matrix $\mathbf{A}$ is calculated using 3 matrices: 1) $\mathbf{\overline{M}}$, a matrix derived from the graph incidence matrix representing the structural mapping from power flows to the states, 2) a weighted incidence matrix corresponding to convective power flow, and 3) $\mathbf{C}$. Matrix $\mathbf{B}_1$ is obtained from $\mathbf{\overline{M}}$ and $\mathbf{C}$. $\mathbf{B}_2$ is obtained from a weighted incidence matrix corresponding to advective and bidirectional advective power flows. For a detailed explanation of these matrices, please refer to reference ~\cite{peddada2019optimal}. 
\small{
\begin{align}
\label{Eq: Dynamics}
    \dot{\bm T}(t)=\bm A \begin{bmatrix} 
                        \bm T(t)\\
                        T^t(t)
                       \end{bmatrix}+
                       \bm B_1 \left(diag\left( \bm Z\begin{bmatrix}
                       \dot{m}_p(t)\\
                       \dot{\bm m}_f(t)\\
                       \dot{m}_t(t)
                        \end{bmatrix} \right)\right)\bm B_2\begin{bmatrix} 
                        \bm T(t)\\
                        T^t(t)
                       \end{bmatrix}+\bm C^{-1}\bm D \bm P^s(t)
\end{align}}

\section{Generating Multi-split Spatial Graphs}
\label{Sec: Generating Multi-split Spatial Graphs}
\subsection{Graph Representation}
\label{Sec: Representation Multi-split Spatial Graphs}
\begin{figure}[ht!]
    \centering
    \subcaptionbox{$0\, (1,2) (3)$}{\includegraphics[width=0.17\linewidth]{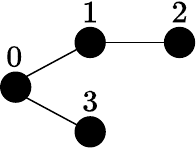}}
    \hspace{1.2 cm}
    \subcaptionbox{$0 \left(1, 2 (3,4) (5)\right) (7)  \left(8 (9,10)\right) (11)$}{\includegraphics[width=0.32\linewidth]{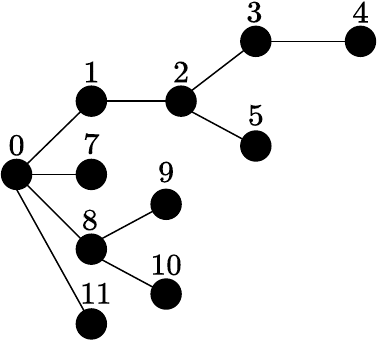}}
    \caption{Two multi-split representations and their equivalent graphs.}
    \label{fig: twoGraphReps}
\end{figure}

The system configurations here can be represented by acyclic undirected connected graphs where any two nodes are connected by only one simple path ( i.e. a tree). In the current representation, the tank is always the root node and each CPHX is labeled with a number from 1 to N. Splittings occur at the root or at any other node. In the corresponding symbolic representation, branching is shown by parentheses and consequent CPHXs, i.e., nodes in one branch are separated by a comma. Figure~\ref{fig: twoGraphReps} shows the representation for two configurations and their equivalent graphs.   

\begin{figure}[ht!]
    \centering
    \subcaptionbox{}{\includegraphics[width=0.49\linewidth]{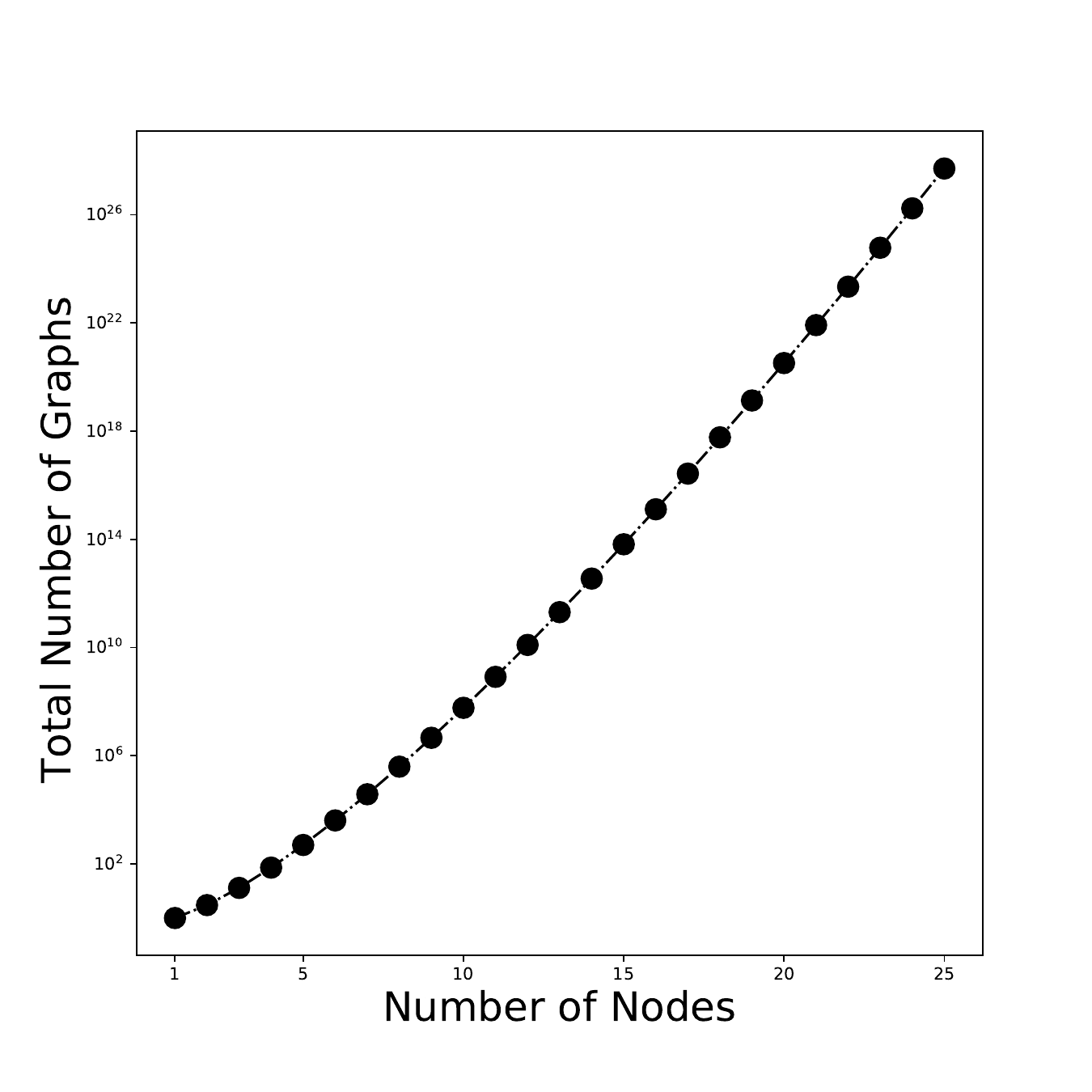}} 
    \subcaptionbox{}{\includegraphics[width=0.49\linewidth]{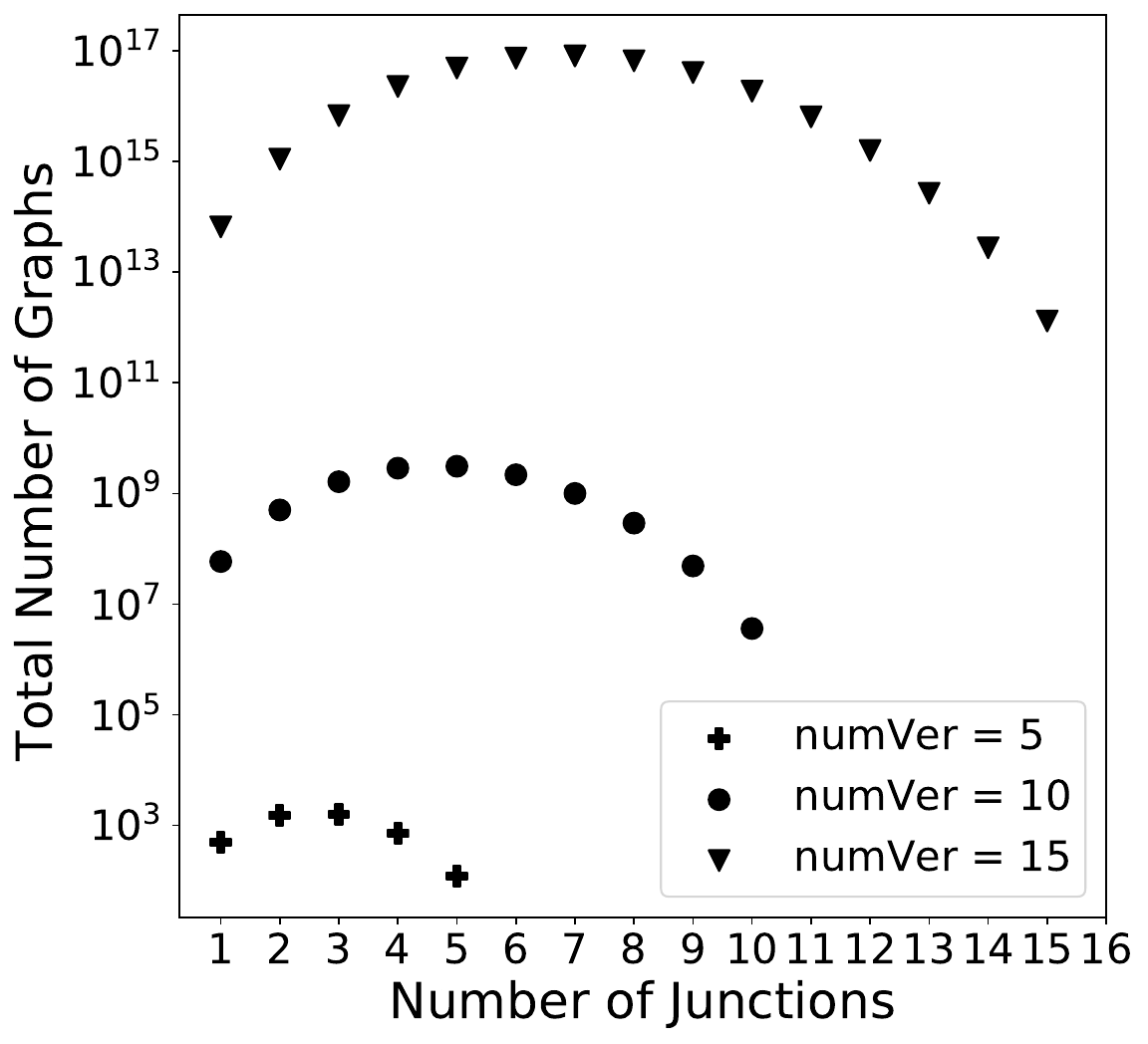}} 
    \subcaptionbox{}{\includegraphics[width=0.49\linewidth]{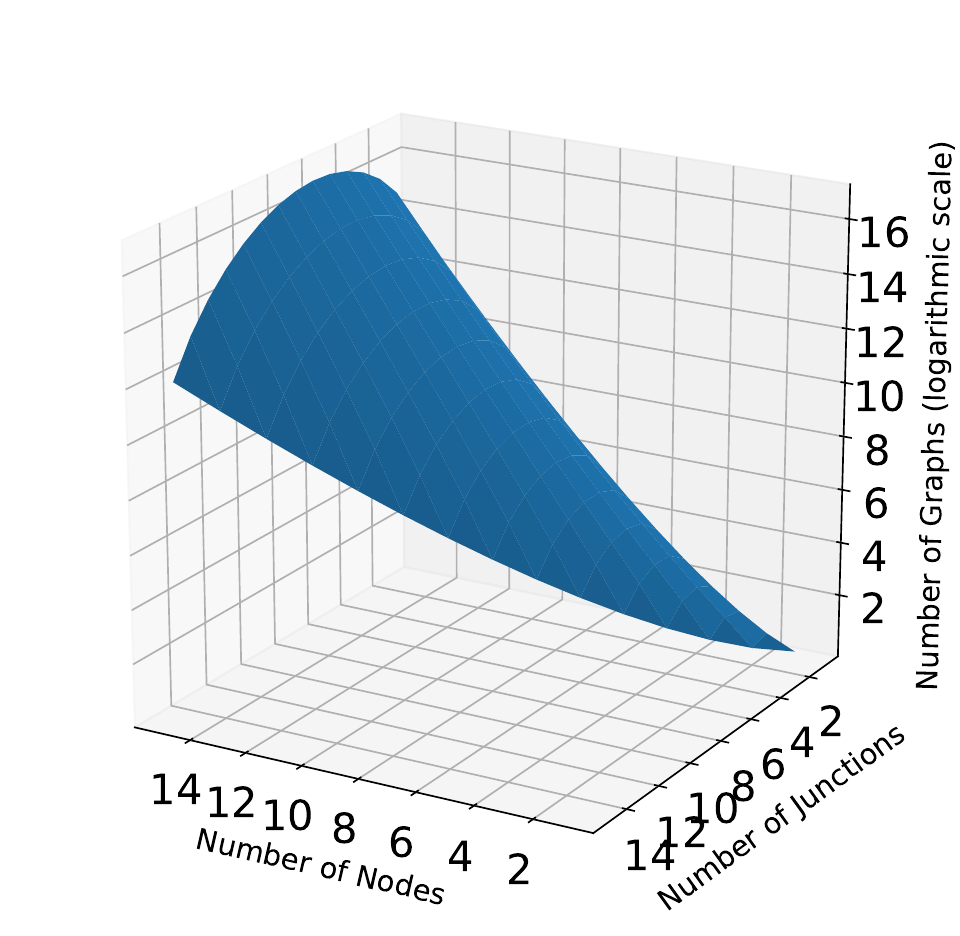}}
    \subcaptionbox{}{\includegraphics[width=0.49\linewidth]{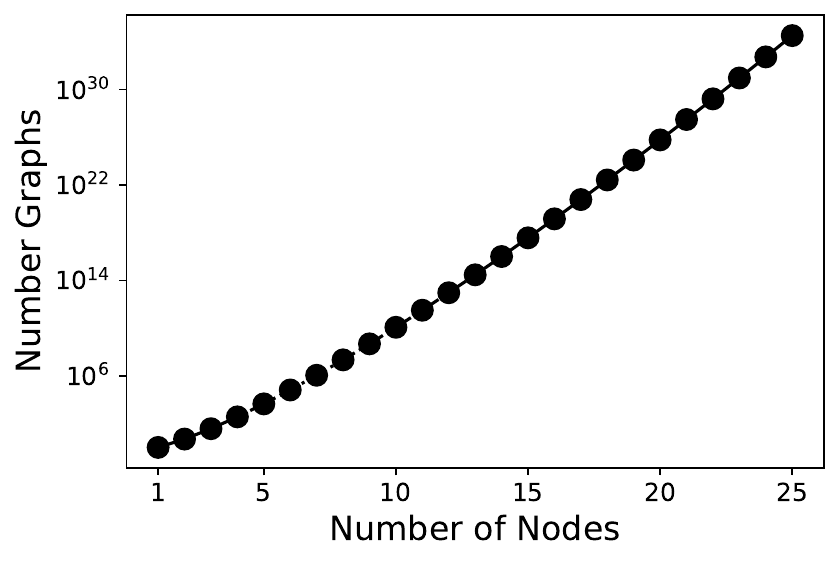}}
    \caption{
    (a) Number of single-split graphs vs. number of non-root nodes, (b) The number of multi-split graphs for graphs with $N=$ 5, 10, and 15 nodes when the number of junctions varies from 1 to $N$, (c) 3D surface illustrating the number of graphs as a function of the number of vertices and junctions, and (d) the total number of multi-split graphs with only one junction layer as a function of $N$.}
    \label{fig: num_oneLayer_multiSplit_graphs}
\end{figure}

The number of single-split configurations increases rapidly as the number of heat-generating devices increases. Figure~\ref{fig: num_oneLayer_multiSplit_graphs}(a) shows the total number of single-split configurations as a function of number of nodes in graphs. Equation~\eqref{Eq: Number of Single-Split Graphs} quantifies the number of generated configurations for this enumeration. When splits (or \textit{junctions}) are added, the total number of configurations increases more. Assuming there is only one-layer of junction nodes, the number of configurations depends on the number of junctions as well as the number of non-junction nodes. One can calculate the number of configurations for a fixed number of non-junction nodes ($N$) and a variable number of junctions, ranging from 1 to $J$, in a recursive manner. Table~\ref{tab:numGraphWithJnks} summarizes the equations.

\begin{align}
\label{Eq: Number of Single-Split Graphs}
    G(N) = \sum_{k=0}^{N} {N\choose k} {N-1\choose k-1} (N-k)!
\end{align}

\begin{table}[ht!]
\small
\centering
\caption{Number of graphs for a given number of junctions}
\label{tab:numGraphWithJnks}       
\begin{tabular}{cc}
\toprule
Junctions     & Number of Graphs   \\ \hline
J = 1        & $F_1(N) = G(N)$ \\
J = 2        & $F_2(N) = \sum_{M=1}^{N} {N\choose M} G(M) F_1(N-M)$ \\[2pt]
J = J & $F_J(N) = \sum_{M=1}^{N} {N\choose M} G(M) F_{J-1}(N-M)$ \\  \bottomrule
\end{tabular}
\end{table}

Figure~\ref{fig: num_oneLayer_multiSplit_graphs}(b) presents the number of configurations generated for graphs with N = 5, 10, and 15 when J varies from 1 to N. Figure~\ref{fig: num_oneLayer_multiSplit_graphs}(c) shows the number of graphs as a function of N and J. To find the total number of graphs with one junction layer (J = 1 to N) , we need to sum up all the graphs generated for different number of junctions; see Fig.~\ref{fig: num_oneLayer_multiSplit_graphs}(d) for an illustration of this growth.

The addition of junction nodes and layers leads to a rapid growth in the number of graphs. It is desirable to impose constraints in a way that meaningfully limits this growth. One strategy is to recognize that these graphs represent physical designs, and that spatial system information can be leveraged to reduce problem complexity. Here we employ a spatial clustering method to group the nodes in sub-domains and enumerate the sub-branches only in the nearby neighborhoods; this procedure is detailed in Section ~\ref{Sec: Generation Multi-split Spatial Graphs}. This approach, however, cannot access all possible designs (as can enumeration). Therefore, we examine a second strategy where the junction locations are enumerated. Exploring the trade-offs between performance improvement and reducing problem complexity is explored in this article as in some cases enumeration may be worth the added computational cost. 

\subsection{Generation of Multi-split Spatial Graphs}
\label{Sec: Generation Multi-split Spatial Graphs}
Algorithms \ref{Clustering-Alg}--~\ref{Graph-Alg} describe the multi-split graph generation framework of this work. The generation procedure passively choose junction nodes during run-time instead of employing all the nodes in enumerations from the beginning. It performs recursion for refinement. The enumerations can cover (i) tank-junctions, (ii) junction-CPHXs, and (iii) both, based on the specific requirements of the optimization problem; this study focuses on class (ii).

\begin{figure}[ht!]
    \centering
    {\includegraphics[width=1\linewidth]{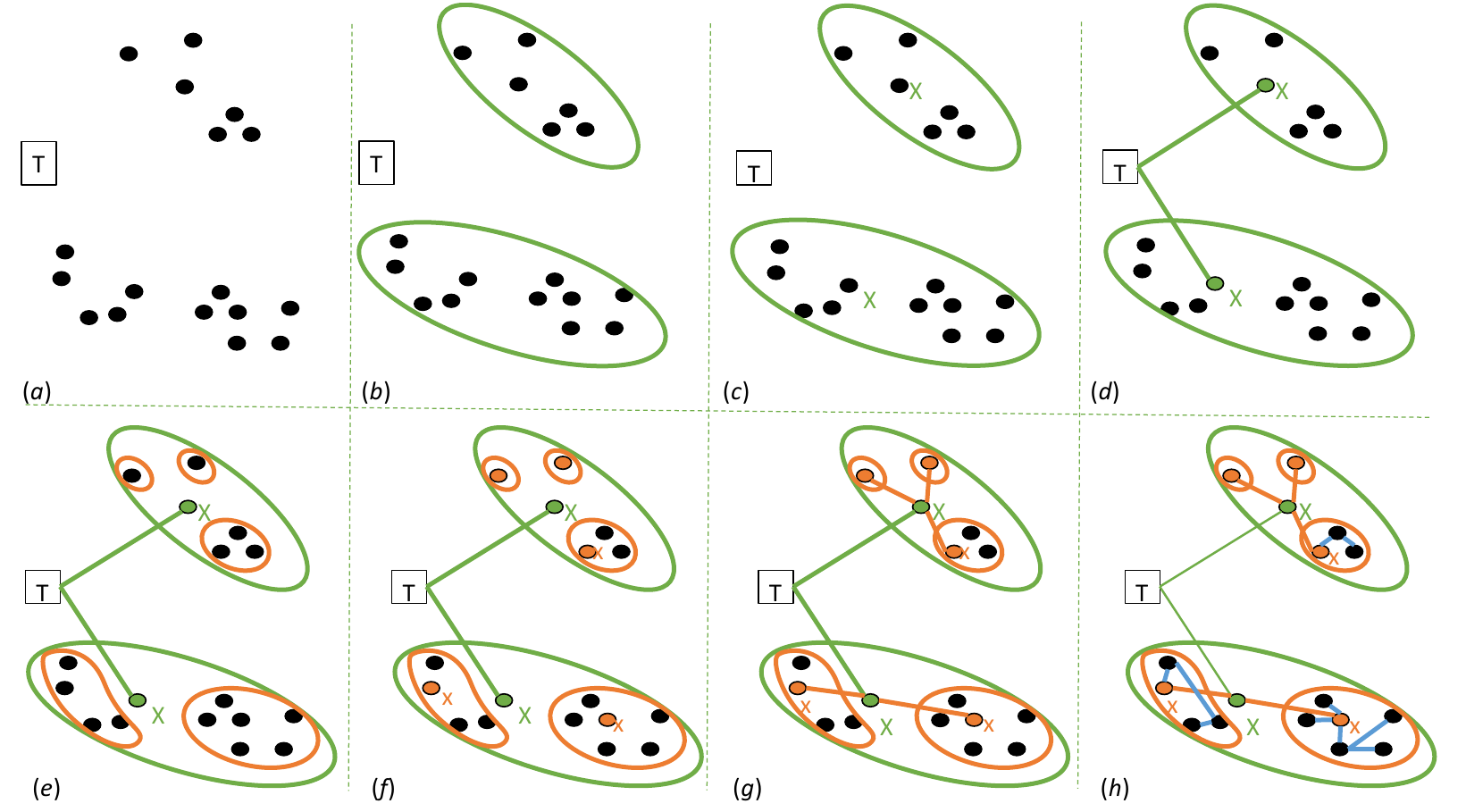}} 
    {\includegraphics[width=0.3\linewidth]{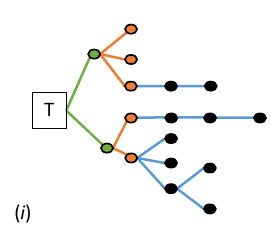}} 
    \caption{Clustering and the creation of a sample configuration for a set of nodes. Figure shows (a) the spatial locations of source (T) and 17 nodes, (b) first-level clusters, (c) centroid of each cluster, (d) chosen junctions and a routing from T to junctions, (e) second-level clusters, (f) the centroid and the corresponding junctions of the second-level, (g) a routing from first-level to second-level junctions, (h) a sample enumeration in the internal cluster with more than 1 node. Figure (i) presents the generated sample graph. }
    \label{fig:clusteringGraphRouting}
\end{figure}

\begin{algorithm}[ht!]
\footnotesize
\caption{Selecting Number of Clusters to form Super-Nodes}\label{Clustering-Alg}
\begin{algorithmic}[1]
\State $N \gets $ number of nodes
\For{$K$ in range(1, $N$)}
  \State cluster nodes into $K$ clusters via K-means
  \State if the clustering is stable, stop
\EndFor
\State return $K$ and the corresponding nodes of each cluster
\end{algorithmic}
\end{algorithm}

\begin{algorithm}[ht!]
  \caption{Generation of Hierarchical Tree of Super-Nodes }\label{Tree-Alg}
  \footnotesize
  \begin{algorithmic}[1]
    \State $T\gets$ tree
    \State super-node[0]$\gets$ $ \{P[0], D[0]\}$ here $P[0]$ is Tanks and $D[0]$ is the spatial data of all nodes 
    \State $L[0]$ includes super-node[0]
    \State $numL\gets$ number of levels in the tree
    \State $m\gets$ number of super-nodes in $T$ 
    \For{$l $ in range(1, $numL$)} 
    \State $Ids\gets$ index of all super-nodes in $L[l-1]$ 
    \For{$i$ in $Ids$} 
    \State cluster super-node $i$ into $K[i]$ clusters using Algorithm ~\ref{Clustering-Alg} 
    \For{$k $ in range(1,K[i])} 
    \State  $ m = m+1 $
    \State  $P_k \gets $ the closest node to centroid of $D(k)$ ($P_k$ is  junction.)
    \State $ P[i] \gets $ $[P_1, P_2, ..., P_n,P_k]$ 
    \State remove $P_k$ from $D(k)$ 
    \State $T[m] \gets $ $ \{P[k], D[k]\}$
    \Stateh $L[l] \gets$ add $T[m]$ 
    \EndFor
    \EndFor
    \EndFor
  \State return Tree T with numL levels
  \end{algorithmic}
\end{algorithm}

\begin{algorithm}[ht!]
\footnotesize
\caption{Graph Enumeration}\label{enumeration-Alg}
\begin{algorithmic}[1]
\State $N \gets$ number of nodes
\State $eG[0] \gets [[0]]$ \Comment each sub-list represents the adjacency list of a graph
\State $eG[1] \gets [[(0,1)]]$
\For{$n$ in [2, $N$]}
  \State $Parents[n] \gets [1, \ldots, n-1]$
  \For{$g$ in $eG[n-1]$}
    \For{each $Node$ in $g$}
      \If{$Node$ belongs to $Parents[n]$}
        \State add edge ($Node$, $n$) to $g$ and add the adjacency list of the new \hspace*{5.6em} graph to $eG[n]$
      \EndIf
    \EndFor
  \EndFor
\EndFor
\State return all enumerated graphs $eG[N]$
\end{algorithmic}
\end{algorithm}

\begin{algorithm}[ht!]
  \caption{Graph Generation for a Selected Tree Level }\label{Graph-Alg}
  \footnotesize
  \begin{algorithmic}[1]
  \State $l \gets $ Tree level selected for graph generation
  \State $Ids \gets $ index of all super-nodes in $L[l]$ for tree generated by Algorithm ~\ref{Tree-Alg}
  \For{$i$ in $Ids$}
  \State $Q \gets$ parents of the super-node ($P[i]$) 
  \State $g1[1:j]$ $\gets$ enumerate super-node$[i]$ using Algorithm ~\ref{enumeration-Alg} with  its junction \hspace*{7.0em} as its root
  \State $g2$ $\gets$ the circular graph of $Q$ with $(P[-1], P[0])$ edge removed
  \State $g[i,1:j]$ $\gets$ merge $g2$ with every\hspace*{0.0em} $g1[1:j]$
  \EndFor
  \State combine a sub-graph selected from each super-nodes of level l to generate all the graphs
  \end{algorithmic}
\end{algorithm}

Algorithm ~\ref{Clustering-Alg} initially creates trees using nodes spatial data. The nodes data are clustered to form \textit{super-nodes} in a recursive manner. Here, a level represents the levels of splitting in an architecture tree. A junction node is defined for each super-node by choosing the node with the smallest Euclidean distance from the cluster's centroid, see Figure ~\ref{fig:clusteringGraphRouting} and Algorithm ~\ref{Tree-Alg}. Note that a junction represents a CPHX node where the coolant flow splits into branches. Next, we enumerate all sub-graphs of a selected tree-level. This means, for a chosen super-node in that tree level, its nodes are enumerated to form all possible sub-graphs; Algorithm ~\ref{Graph-Alg} describes how the enumeration algorithm works. Afterwards, we connect the root node (Tank) and the corresponding junction nodes of the super-node to its sub-graphs, see Algorithm~\ref{enumeration-Alg}. This process is performed for all super-nodes of the selected level. Finally, an architecture graph is created by choosing and merging one sub-graph from the pool of sub-graphs generated for each super-node in that level, refer to Algorithm ~\ref{Graph-Alg}. Note that various sub-clusters can be defined within a cluster to form multiple layers of junction nodes. 

In addition to the above mentioned graph generation algorithm, we also generated graphs using a second strategy where the locations of junctions in the trees are enumerated. The main difference between the first and second graph generation strategies is in topological locations of junction nodes in the architectural graphs.

Prior research performed by some of the authors on holistic design for 3D spatial packaging and routing of interconnected systems~\cite{Peddada2020, Peddada2020_JMD_2Stage, Peddada2022a, Peddada2020b, Peddada2021} will be employed in future work to capture the spatial aspects of the multi-split configurations via optimal placement of junction nodes and cooling circuit components such as the CPHXs, pumps, valves, and the tank with simultaneous optimization of 3D lengths of the branches (or pipe segments) while satisfying volume and multi-physics constraints. This can lead to a more complete definition of the real-world multi-split configuration fluid-based thermal system problem.

\begin{figure}[ht!]
    \centering
    \includegraphics[width=1.0\linewidth]{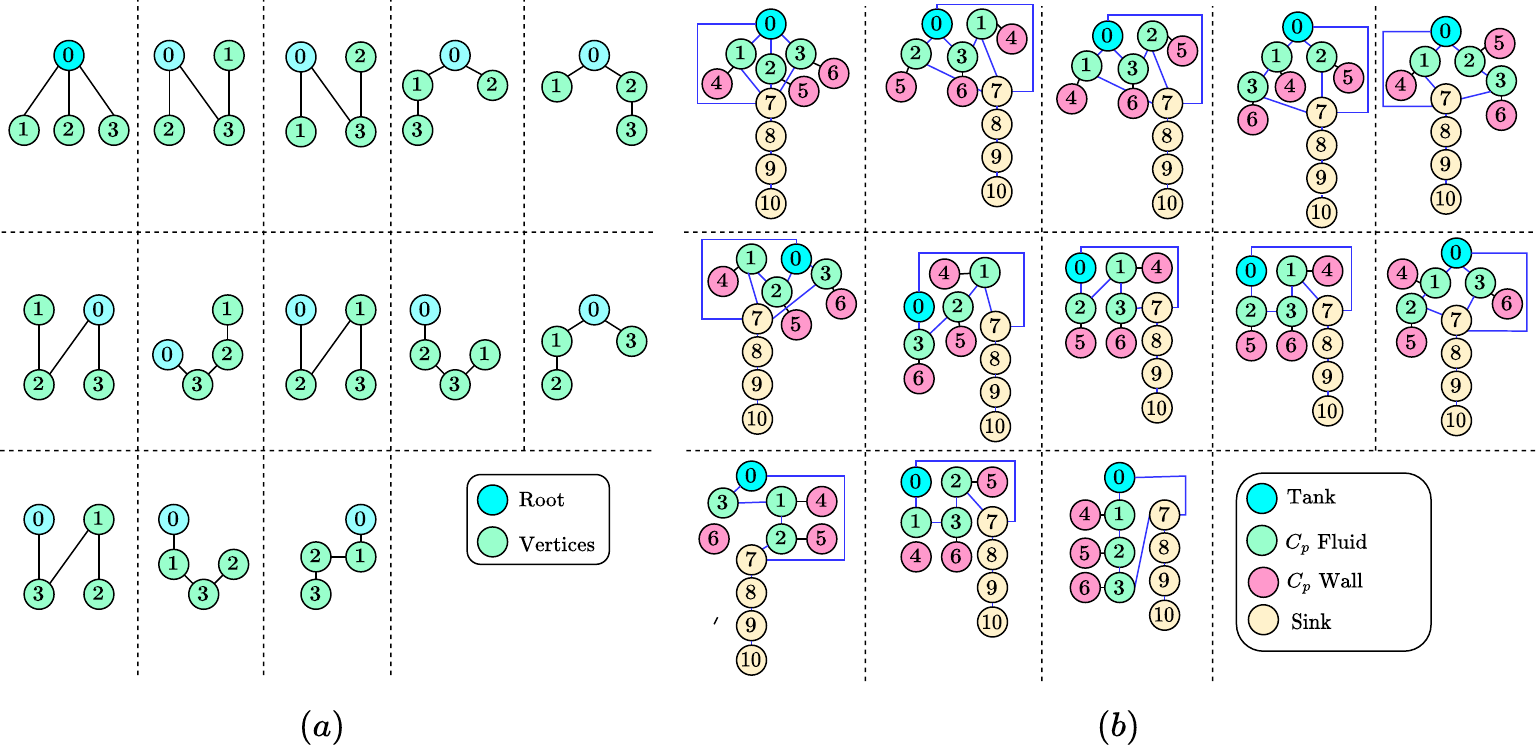}
    \caption{(a) Base and (b) complete physics graphs for single-split systems with 3 CPHXs}
    \label{fig:Base_Complt_Graphs_HXs}
\end{figure}

\begin{figure}[ht!]
    \centering
    \includegraphics[width=0.9\linewidth]{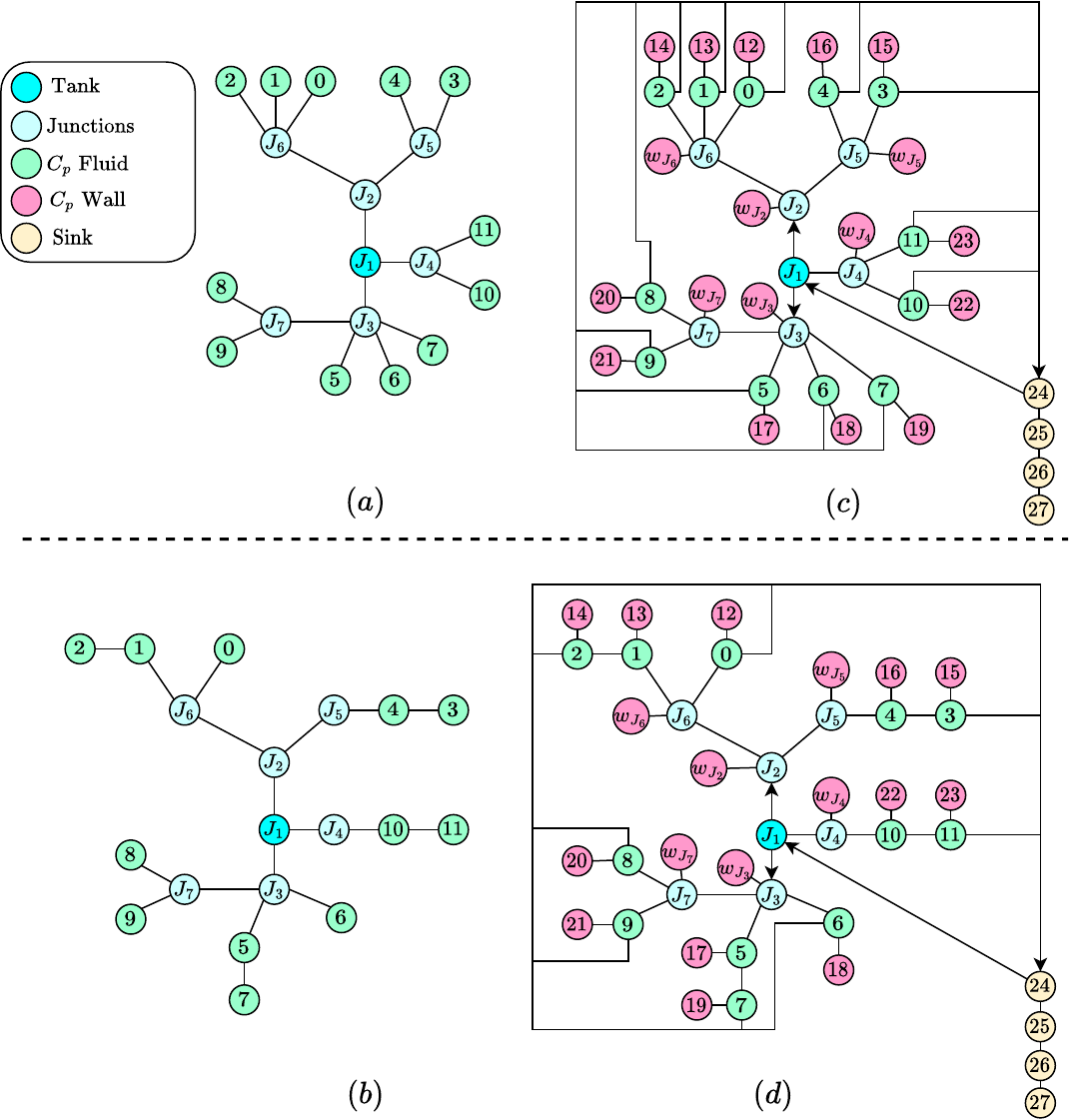}
    \caption{Two sample configurations for a multi-split system with 18 CPHXs. Figures (a) and (b) show the base graphs, and Figs.~(c) and (d) show the complete physics graphs for the base graphs (a) and (b), respectively.}
    \label{fig:configs_19vertices_multiSplit_graphs}
\end{figure}

We generate all configurations within a class, evaluate the performance of each configuration, and choose the optimal one. Figure~\ref{fig:Base_Complt_Graphs_HXs} shows all 13 configurations made by the single-split algorithm for 3 nodes. The figure depicts the base graphs and the extended physics graphs generated for simulations. Figure~\ref{fig:configs_19vertices_multiSplit_graphs} shows two examples configurations and their physics models for a multi-split system with 19 nodes (root, 6 junction CPHXs, and 12 CPHXs nodes). 

\section{Optimal Flow Control Problem (OLOC)}
\label{Sec: Optimal Flow Control Problem}
\begin{figure}[ht!]
    \centering
    \includegraphics[width=1.0\linewidth]{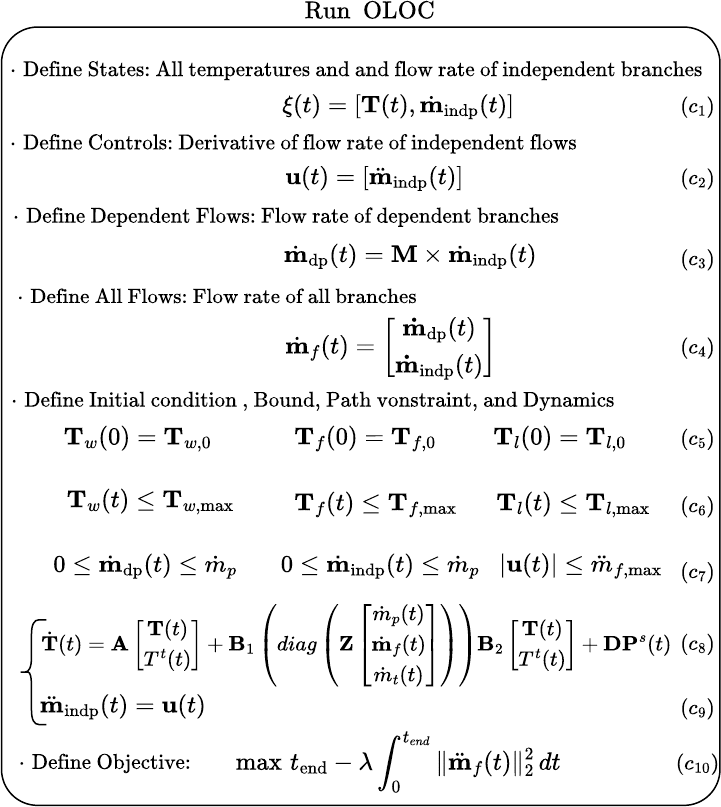}
    \caption{OLOC components}
    \label{fig: OLOC structure}
\end{figure}

The objective of the optimal control problem is to maximize the thermal endurance while satisfying limits on temperatures and mass flow rates. When any of the node's temperatures (defined as states) reaches the upper bound, the OLOC terminates, and the final time is recorded as the thermal endurance. Figure~\ref{fig: OLOC structure} shows the OLOC structure. Here, we seek to determine a control trajectory $u(t)$ for each independent flow that maximizes the objective function while satisfying constraints.
Within this figure, Eq.~($c_1$) presents the system states ($\bm{\xi}$), encompassing the vector of temperature nodes ($\bm{T}$), and the flow rate of independent branches ($\bm{\dot{m}}_{\mathrm{indp}}$). The dynamics that shows how these states evolve over time are shown in Eq.~($c_8$). Equation~($c_2$) defines the control signals as the rate of change of the valve flow rates in independent branches. The flow rate in dependent branches is determined by an algebraic equation that ensures input flow rate of each branch is equal to the output flow rate of that branch. This equation can be represented as a matrix multiplication, as shown in Eq.~($c_3$), where matrix $M$ maps the flow rate of independent branches ($\bm{\dot{m}}_{\mathrm{indp}}$) to the flow rate of dependent branches ($\bm{\dot{m}}_{\mathrm{dp}}$). A practical illustration of this calculation is provided in Figure \ref{Fig:Graph Signal}. In Eq.~($c_4$), the total flow rate of all branches is presented, comprising both independent and dependent flow rates. It is worth mentioning that the order of combining independent and dependent flow rates depends on the graph structure. For simplicity, in this case, we first incorporate the independent flow rates and then the dependent flow rates.
The initial conditions for the temperature nodes are given by Eq.~($c_5$). In this equation, $\bm{T}_w$ represents the wall temperature of the CPHX states, $\bm{T}_f$ represents the fluid temperature of the CPHX states, and $\bm{T}_l$ represents the temperature of the Tank and LLHX states.
Eq.~($c_6$) presents inequality path constraints that ensure the operating temperature of each component remains within an upper bound throughout the entire time horizon. The first two terms in Eq.~($c_7$) represent the inequality path constraint that guarantees the flow rates of both independent and dependent branches remain within a specified bound. This equation, when combined with Eq.~($c_3$), ensures that the input flow rate of each branch is equal to its output flow rate. It also ensures that the maximum flow rate cannot exceed the pump flow rate ($\dot{m}_p$). The last term in Eq.~($c_7$) represents the limit on the derivative of the flow rate, capturing the physical limitations of the valves \cite{peddada2019optimal}. Equations $c_8$ and $c_9$ show how states defined in Eq.~($c_1$) evolve over time. In Equation~($c_{10}$), the objective value is represented, aiming to maximize the thermal endurance. Similar to the approach in Ref.~\cite{peddada2019optimal}, a penalty term is incorporated to facilitate solution smoothness and enhance convergence. The parameter $\lambda$ is selected such that the total penalty cost remains below $1\%$ of $t_{\mathrm{end}}$. Table \ref{tab: Problem parameters} shows the parameters used in the physical simulations for the studies in this article.

\begin{table}[ht!]
\small
\centering
\caption{Parameters used in the physics modeling of the thermal systems \cite{peddada2019optimal}.}
\label{tab: Problem parameters}
\scalebox{1.0}{
\begin{tabular}{rl}
\toprule
Parameter                           & Value      \\ \hline
LLHX wall mass                      & 1.2 kg     \\
CPHX wall mass                      & 1.15 kg    \\
Tank fluid mass                     & 2.01 kg    \\
Thermal sink temperature $T^t$         & $15^o$ C       \\
Tank/LLHX initial temperature , $T_{l,0}$ & $15^o$ C       \\
CPHX initall wall temperature, $T_{w,0}$  & $20^o$ C       \\
CPHX inital fluid temperatue , $T_{f,0}$  & $20^o$ C       \\
Thermal sink mass flow rate, $\dot{m}_t$ & 0.2 kg/s   \\
Pupmp mass flow rate $\dot{m}_p$          & 0.4 kg/s   \\
Valve rate limit $\ddot{m}_{f,\mathrm{max}}$              & 0.05 kg/$\mathrm{s}^2$ \\
Penalty parameter, $\lambda$  & $0.01/(N_f {\ddot{\bm{m}}}_{\mathrm{f,max}}^2)$ \\\bottomrule
\end{tabular}}
\end{table}

There are two main approaches for solving OLOC problems: 1) Indirect (optimize then discretize) and 2) Direct (discretize then optimize). In the indirect method, a differential algebraic equation is derived using optimality conditions (the calculus of variations or the Pontryagin minimum principle). These equations should then be discretized and solved numerically \cite{bayat2023ss}. In contrast, the direct method first discretizes the problem so it can be transformed into a nonlinear program (NLP), which can then be solved by a nonlinear programming solver such as SNOPT \cite{gill2005snopt} or IPOPT \cite{biegler2009large}. The indirect method provides more information about the structure of the problem, but solving constrained problem can be challenging with this method. Alternatively, direct methods use NLP solvers to solve complex problems successfully; some well-established OLOC software tools based on the direct method are available, for example GPOPS \cite{patterson2014gpops} and Dymos \cite{falck2021dymos}. This paper uses Dymos to solve the OLOC problems, which is an open source program developed in Python. The computation cost of solving the OLOC problem depends on the size of the graph, but on average, it takes approximately two minutes to solve the problem for each configuration. Additionally, it is important to highlight that in this article, the evaluation of the nonlinear optimal control problems for each of the architectures has been parallelized. This parallelization approach significantly reduces the computational cost associated with solving these problems. All reported computational costs were obtained using a workstation with an AMD EPYC 7502 32-Core Processor @ 2.5 GHz, 64 GB DDR4-3200 RAM, LINUX Ubuntu 20.04.1, and Python 3.8.10.

\section{Case Studies}
\label{Studies}
We present three case studies to illustrate how this work can help engineers design optimal thermal management systems. In the first case study (Sect.~\ref{Sec: brute force of 3 and 4 nodes in both single and multi split}), the goal is to obtain the optimal structure for thermal management systems having 3 and 4 CPHXs. Here we use enumeration and compare both single-split and multi-split cases. For 3 and 4 CPHXs systems, the heat loads are $[12,4,1]$ kW and $[12,4,1,1]$ kW, respectively. In Sect.~\ref{sec: Brute Force of multi with 6 nodes and investigate innerloop result}, we have shown the results for the multiple-split case under two disturbances: $[5,5,5,5,5,5]$ kW and $[5,7,6,4,5,5]$ kW. This example indicates optimal configuration can change when disturbances are different. OLOC signals are also compared for some configurations. While the main focus of this article revolves around the results obtained from small graphs for detailed analysis and discussions, it is important to note that the code is applicable to graphs of any size. This versatility is demonstrated in Sec.~\ref{sec: Multi-split configurations with 17 nodes}, where a graph consisting of 18  nodes, with a Tank, 14 CPHXs, and 3 junctions CPHXs, is studied.

\subsection{Comparing single-split and multi-split cases with 3 and 4 CPHX-nodes}
\label{Sec: brute force of 3 and 4 nodes in both single and multi split}

\begin{figure}[ht!]
    \centering
    \subcaptionbox{Population ranking}{\includegraphics[width=1.0\linewidth]{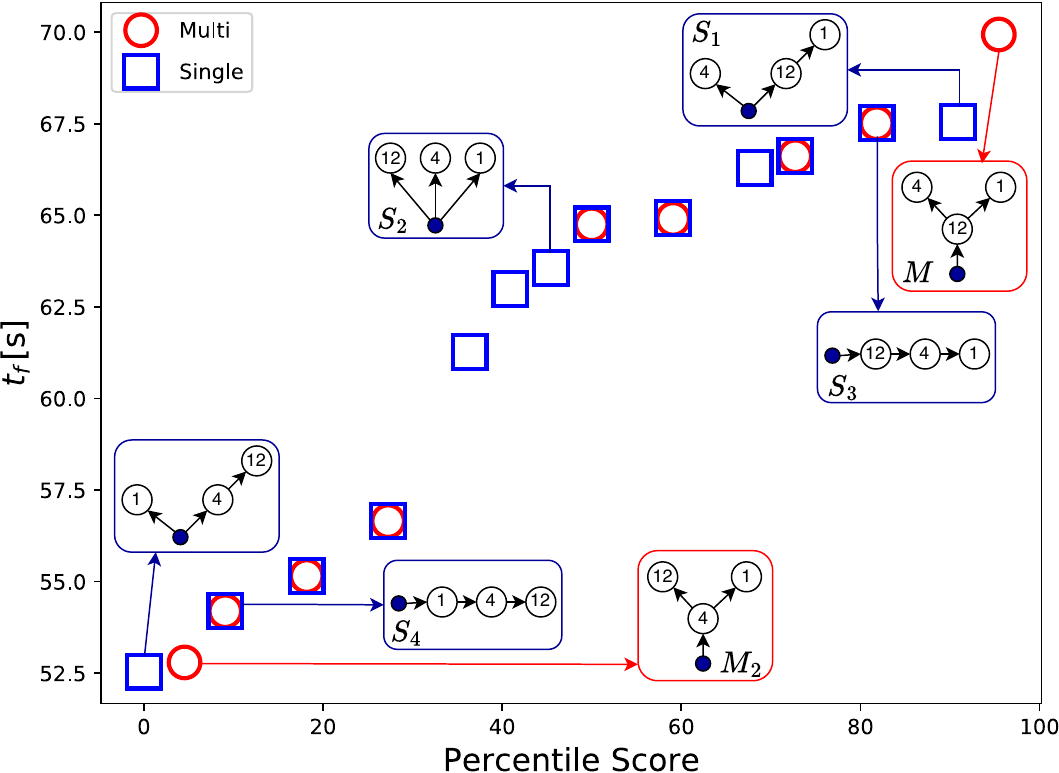}}
    \vskip 10pt
    \subcaptionbox{Flow rate of the cases shown in Fig.\ref{comparison_multi_single_3_and_inner_results}(a)}{\includegraphics[width=1.0\linewidth]{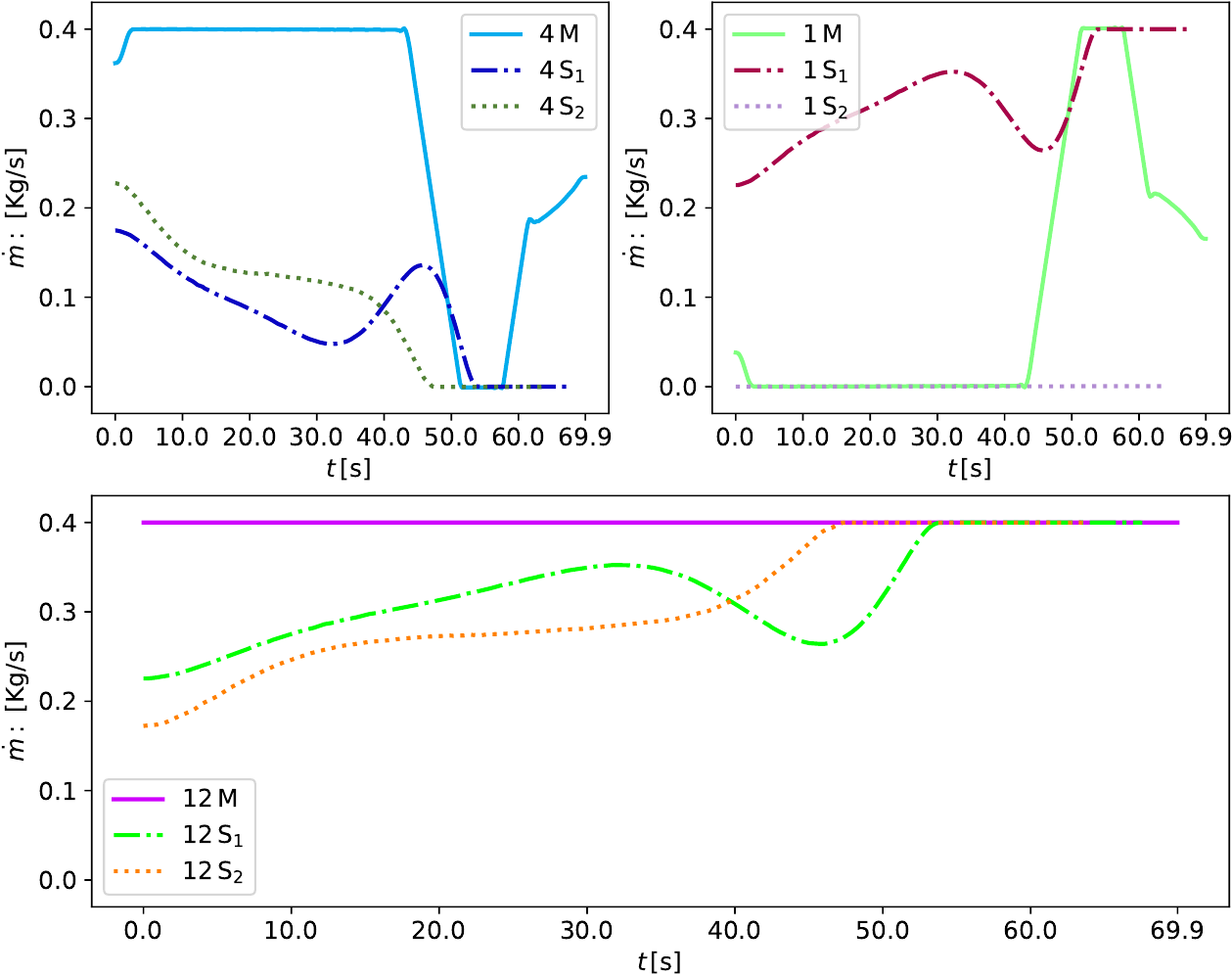}}
    \vskip 10pt
    \subcaptionbox{Wall temperature of the cases shown in Fig.\ref{comparison_multi_single_3_and_inner_results}(a)}{\includegraphics[width=1.0\linewidth]{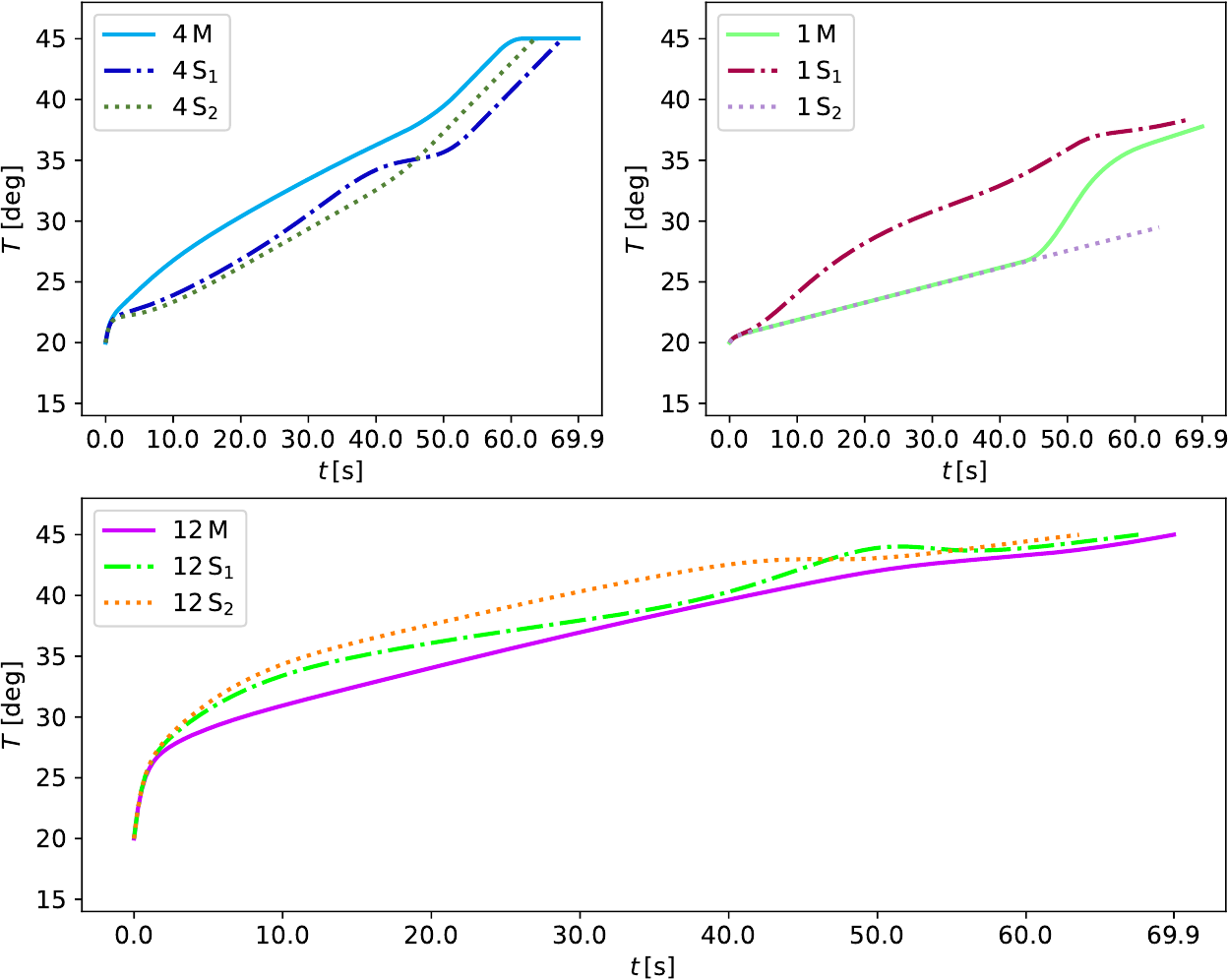}}
    \caption{Comparison of multi-split and single-split cases with 3 nodes having $dist=[12, 4, 1]$ kW.}
    \label{comparison_multi_single_3_and_inner_results}
\end{figure}

Here we aim to find the best architectures among single-split and multi-split configurations with 3 and 4 nodes. The result for the first scenario with 3 nodes is shown in Fig.~\ref{comparison_multi_single_3_and_inner_results}. In this illustration, only the fluid nodes of CPHX and the tank are depicted, while the disturbance of each CPHX is displayed in its respective node. Here, \textit{Multi} represents multi-split cases and \textit{Single} represents single-split cases. The horizontal axis in Fig.~\ref{comparison_multi_single_3_and_inner_results}(a) shows the percentile score and the vertical axis shows thermal endurance. A percentile score represents the relative position of a value within a dataset by indicating the percentage of values that are lower than it. Thus, the best case is at the top right, and the worst case is at the bottom left. As we see, the multi-split architecture yield the best result. It should be mentioned that the results obtained depend on the heat load. For example, here, we have an extreme load (12) that is much larger than other loads. Here, the multi-split configurations usually has a better result since the node with the maximum disturbance is connected to the tank and receives the maximum available flow-rate. However, in many other cases, the flow rate that this node receives is a fraction of the pump flow rate.

In Fig.~\ref{comparison_multi_single_3_and_inner_results}, three cases are denoted as $M$ (Multi-split), $S_1$ (single-split-1), and $S_2$ (single-split-2). The flow rate and wall temperature of these three cases are studies in Fig.~\ref{comparison_multi_single_3_and_inner_results}(b), and Fig.~\ref{comparison_multi_single_3_and_inner_results}(c). Among the three cases, $S_2$ exhibits the highest control authority due to its maximum parallel flows (3). This suggests that the best results can be expected from this configuration. However, this assumption may not necessarily hold true. A significant difference between $M$ and $S_2$ lies in how the node with a 12 kW heat load is handled. In the case of $M$, this node receives the maximum flow rate (pump flow rate) as it is directly connected to the pump. Conversely, in $S_2$, the flow received by this node is a fraction of the pump flow rate. As depicted in Fig.~\ref{comparison_multi_single_3_and_inner_results}(b), the flow rates of these nodes differ across each graph. Consequently, this disparity impacts the objective function value and results in varying temperatures. Additionally, Fig.~\ref{comparison_multi_single_3_and_inner_results}(c) demonstrates that the wall temperature of nodes 12 and 4 reached the upper bounds. Notably, the case labeled as "$M$" achieves this upper bound at a later stage compared to the other cases, indicating a better objective function value (specifically, 69.9 $^{\circ}$ C). 

As mentioned earlier, one of the advantages of using a multi-split graph in this case is that the node with the maximum heat load is directly connected to the pump and receives the maximum flow rate. Therefore, one might expect similar results if all nodes are connected in series directly to the pump. However, as depicted in Fig.~\ref{comparison_multi_single_3_and_inner_results}(a), when these nodes are arranged in series with the pump and nodes with higher loads are positioned closer to the pump ($S_3$), the achieved result, although satisfactory, is inferior to the multi-split case. This disparity arises because the dynamics involved in this problem, such as convection, advection, and bidirectional advection, are also dependent on the graph's structure. Consequently, even though the node 12 in both $M$ and $S_3$ receive the pump flow rate, their distinct dynamics lead to different objective function values. It should also be noted that even with a fixed structure, altering only the load locations yields different outcomes. For instance, as illustrated in Fig.~\ref{comparison_multi_single_3_and_inner_results}(a), cases $S_3$ and $S_4$ possess the same structure but exhibit substantially different objective values ($t_f$). Furthermore, simply changing the positions of the 12 and 4 heat loads results in significantly different objective values for the multiple split cases ($M$ and $M_2$). In these cases, we generally observe that when the structure remains fixed, the objective value tends to be better when nodes with higher heat loads are positioned closer to the pump. This can be attributed to the fact that nodes with high heat loads require cooler fluid to dissipate the heat. When these nodes are closer to the pump, the fluid reaching them is relatively cooler. However, if these nodes are located far away from the pump, the fluid reaching them is already hot as it has absorbed heat from other nodes. Consequently, the objective value (in this case, thermal endurance) decreases. Such studies provide valuable insights for engineering purposes, enabling us to extract knowledge from optimization data. In our future work, we intend to expand upon this idea to extract interpretable knowledge that is understandable to humans.

\begin{figure}[ht!]
    \centering
    \subcaptionbox{Population ranking}{\includegraphics[width=1.0\linewidth]{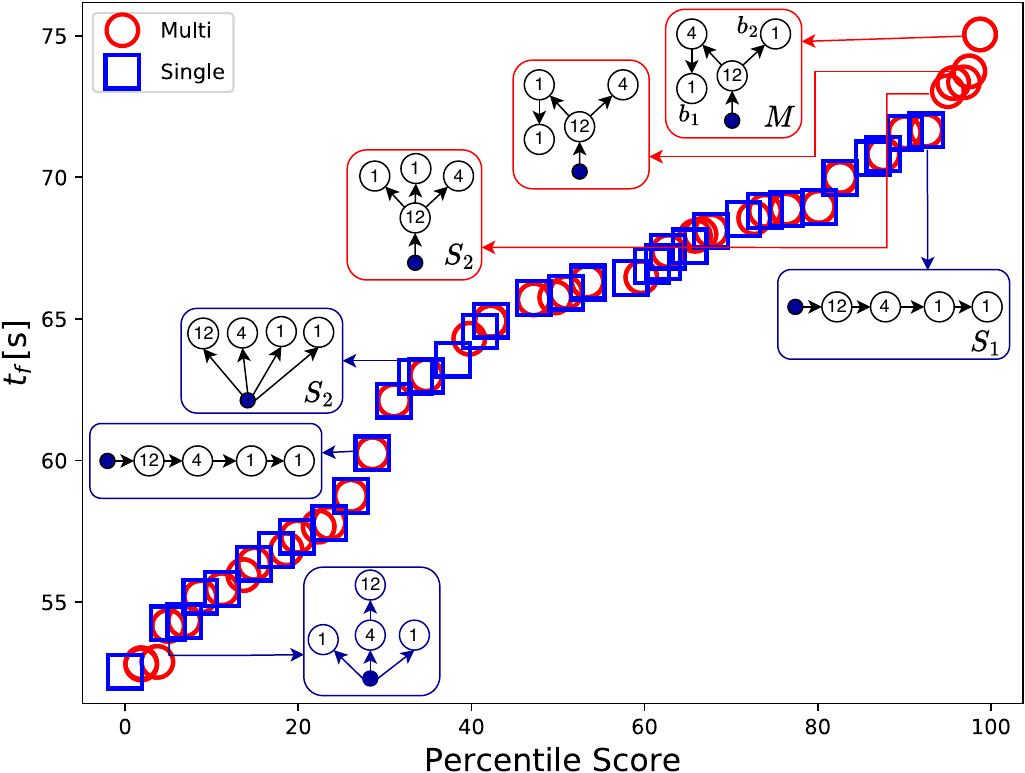}}
    \vskip 10pt
    \subcaptionbox{Flow rate of the cases shown in Fig.\ref{comparison_multi_single_4_and_inner_results}(a)}{\includegraphics[width=1.0\linewidth]{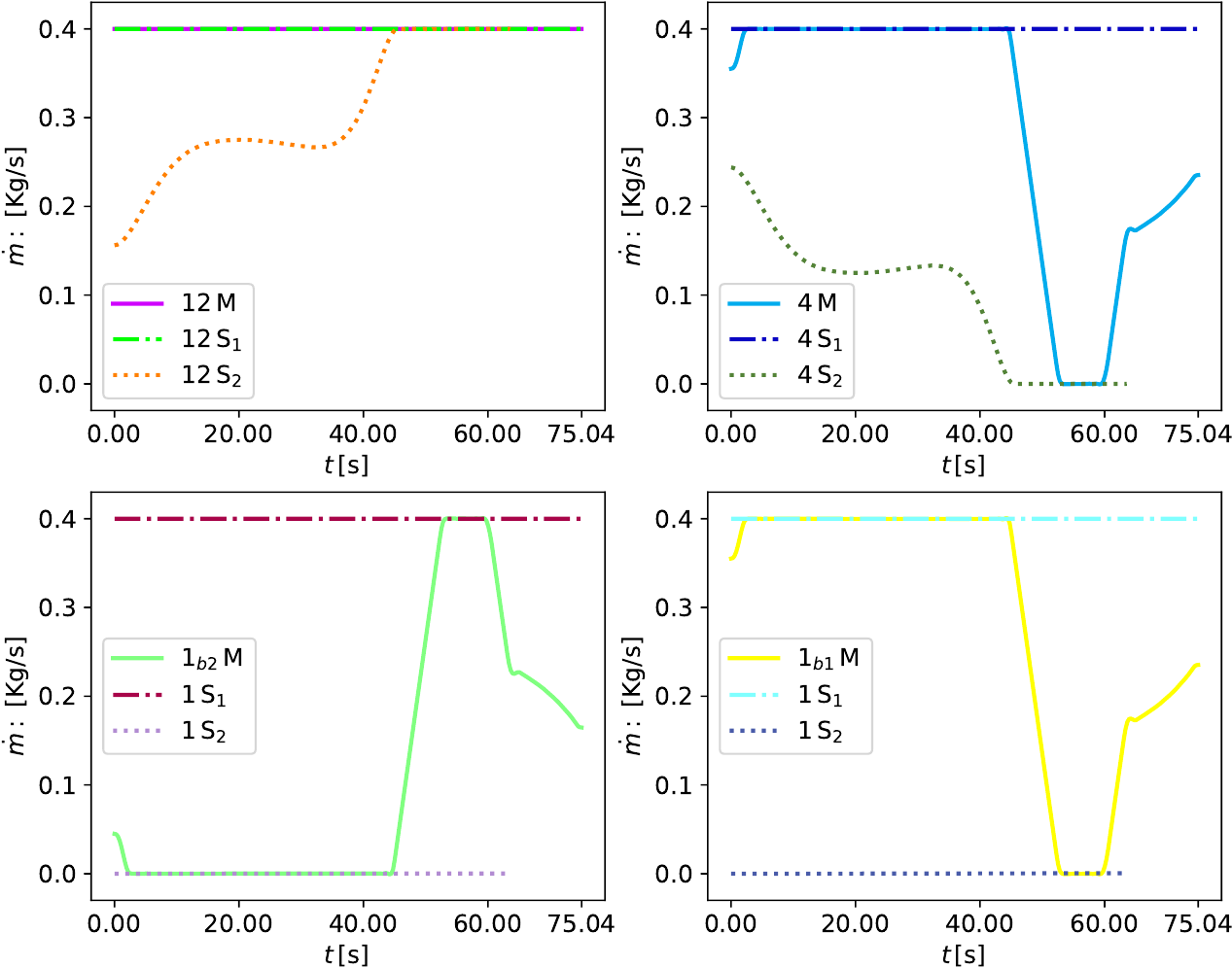}}
    \vskip 10pt
    \subcaptionbox{Wall temperature of the cases shown in Fig.\ref{comparison_multi_single_4_and_inner_results}(a)}{\includegraphics[width=1.0\linewidth]{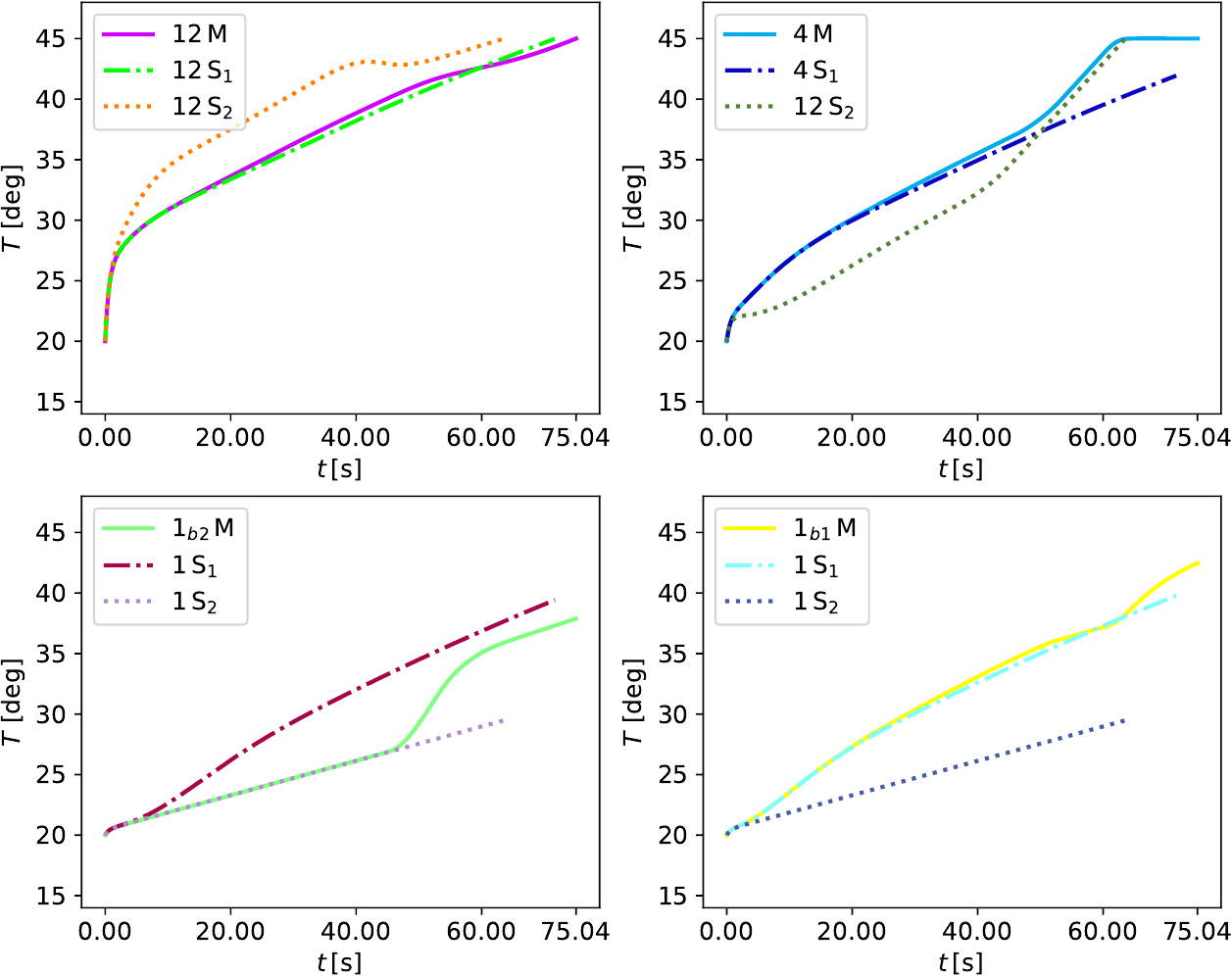}}
    \caption{Comparison of multi-split and single-split cases with 4 nodes, having $dist=[12, 4, 1, 1]$ kW.}
    \label{comparison_multi_single_4_and_inner_results}
\end{figure}

Figure~\ref{comparison_multi_single_4_and_inner_results} presents a similar study, but this time for graphs with 4 CPHXs. In the case where we have two nodes with the same heat load ($1$), denoted as $b_1$ in the first branch and $b_2$ in the second branch of case $M$, we observe that the multi-split case ($M$) achieves a higher objective function value compared to the single-split cases. The optimal solution is found in the multi-split case ($M$), where the node with the maximum load is connected to the pump and then divides into two branches. Among the single-split cases, the best solution is achieved when all nodes are arranged in series and the nodes with higher heat loads are positioned closer to the tank ($S_1$). On the other hand, if the order of heat loads is reversed ($S_3$), the objective value worsens. This is because in this scenario, the node with the highest heat load receives fluid that is already hot since it has absorbed heat from other nodes, resulting in a decrease in thermal endurance.

The flow rate and wall temperature characteristics of three graphs, namely $M$, $S_1$, and $S_2$, are investigated in Figure~\ref{comparison_multi_single_4_and_inner_results}(b) and Figure~\ref{comparison_multi_single_4_and_inner_results}(c). 
One notable difference between $M$ and $S_1$ is that, in $S_1$, the flow rate in all nodes is the same as the pump flow rate, whereas in $M$, it is not. As depicted in Fig.~\ref{comparison_multi_single_4_and_inner_results}, the flow rate in node $1_{b2}$ of $M$ is nearly zero initially, increases, and then decreases again. An interesting observation is the change in the wall temperature of node $1_{b2}$. As shown in Fig.~\ref{comparison_multi_single_4_and_inner_results}(c), the wall temperature of this node increases as the flow rate in that branch increases. This phenomenon can be explained by the fact that node $1_{b2}$ has a lower heat load compared to other nodes, and the coolant fluid entering this node is already warmer than its wall. Consequently, the flow rate in this branch increases to allow the coolant fluid to dissipate some of its heat to the CPHX of this node. Furthermore, as the flow rate in $1_{b2}$ surpasses zero, the coolant flow gets cooler and the rate of change in the wall temperature of node $12$ decreases, resulting in a delayed approach to the upper temperature bound (45 degrees Celsius) and thus increasing thermal endurance. As a result, an optimal coordination of coolant fluid flow rate is achieved, facilitating optimal heat transfer between different nodes and the coolant, ultimately leading to the best objective value. A similar situation was noticed in the case of 3 nodes (see Fig.\ref{comparison_multi_single_3_and_inner_results}) around the 40.0 second mark. At this point, the flow rate of node 1 was increased to help dissipate heat from the fluid flow, resulting in cooler fluid. This, in turn, allowed the other wall nodes to reach the upper bound at a later time, ultimately improving thermal endurance. Gaining such intuitions through human experience alone can be challenging. However, optimization studies like these assist engineers in discovering optimal strategies to solve specific problems. By leveraging these studies, engineers can avoid the need for trial and error, thereby significantly reducing the time and resources required to achieve the desired outcome.

\subsection{Multi-split configurations with 6 CPHX-nodes and investigation of the inner-loop results}
\label{sec: Brute Force of multi with 6 nodes and investigate innerloop result}
\begin{figure}[ht!]
    \centering
    \includegraphics[width=1.0\linewidth]{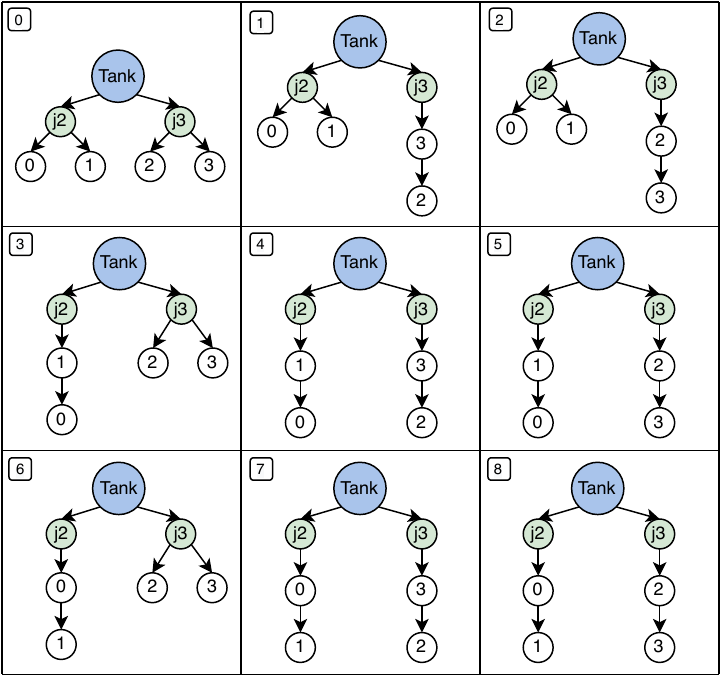}
    \caption{Different architectures produced by the code when the location of CPHXs are defined as: $[[2,0,0],[2,1,0],[3,1,0],[12,12,0],[15,10,0],[13,13,0]]$.}
    \label{fig:investigate_all_configs}
\end{figure}

\begin{figure}[ht!]
    \centering
    \includegraphics[width=1.0\linewidth]{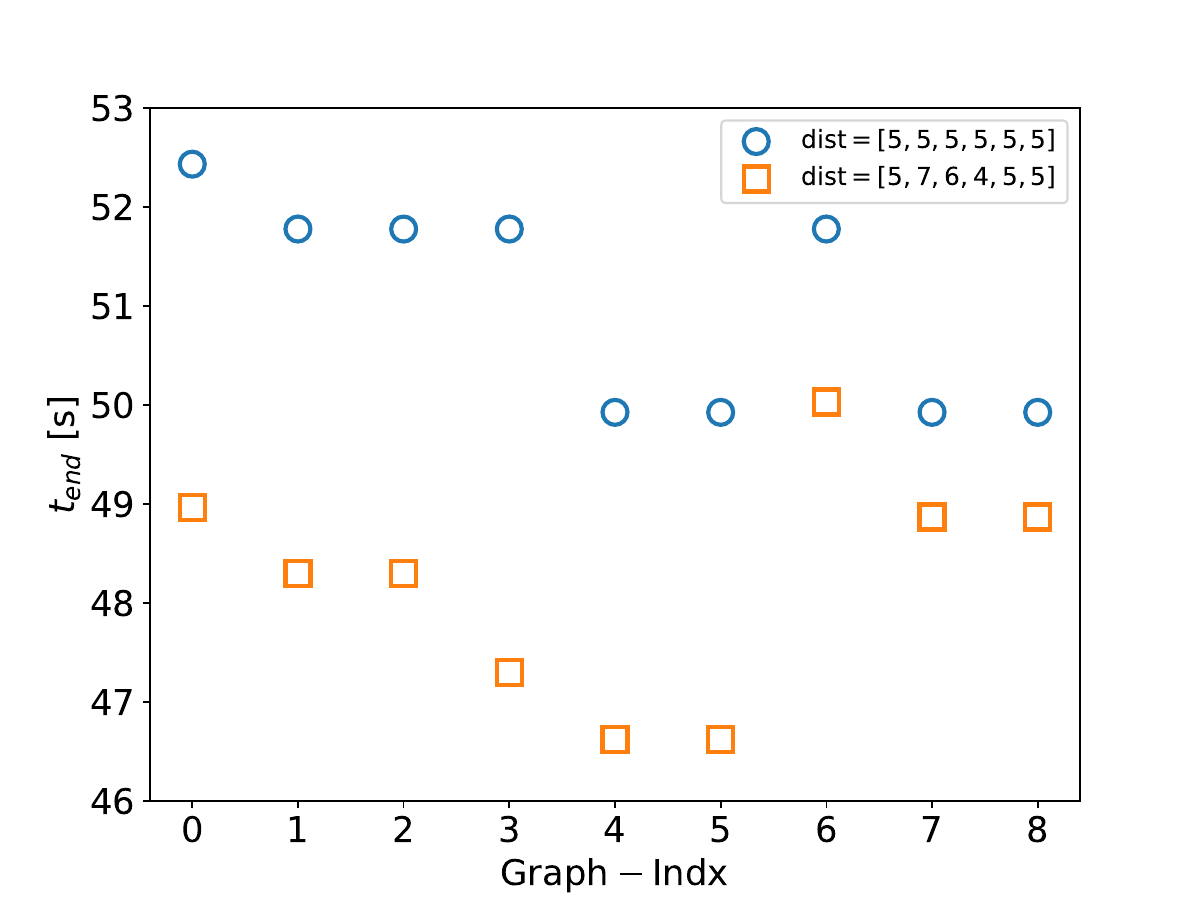}
    \caption{Results obtained for the multi-split configurations shown in Fig. \ref{fig:investigate_all_configs} under two different disturbances. The unit of disturbances is expressed in kW.}
    \label{investigate_brute_force_results}
\end{figure}

In this section we define the locations of the CPHXs, and then, based on spatial location, the junction nodes are produced. For this system, the location of the CPHXs are defined as: $[[2,0,0],[2,1,0],[3,1,0],[12,12,0],[15,10,0],[13,13,0]]$. Two sets of disturbances are considered: case1 = [5,5,5,5,5,5] kW, case2 = [5,7,6,4,5,5] kW. In this structure, the locations of junctions and their heat-load are fixed but all other nodes will vary. Therefore, we have nine different configurations in total, shown in Fig.~\ref{fig:investigate_all_configs}. The result under these two  disturbance sets are shown in Fig.~\ref{investigate_brute_force_results}. As we see, the optimal configuration depends on the disturbance values. For example, for the first scenario, the best result is for the configuration 0, however, for the second scenario, the best result is for configuration 6.

\begin{figure}[ht!]
    \centering
    \includegraphics[width=1.0\linewidth]{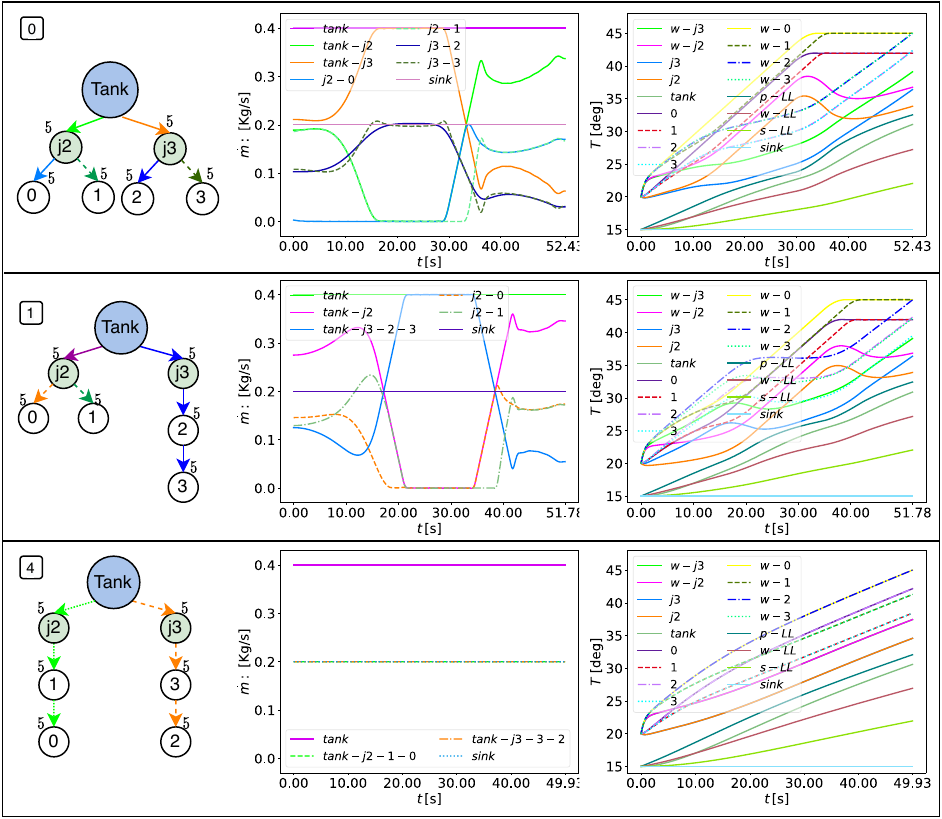}
    \caption{Investigation of the OLOC signals for 3 cases presented in Fig. \ref{fig:investigate_all_configs} and with heat loads = [5,5,5,5,5,5] kW. The multi-split configuration: 0 has the maximum, 4 has the minimum, and 1 has a value in the mid-range of thermal endurance.}
    \label{fig:innerloo_invest_5}
\end{figure}

\begin{figure}[ht!]
    \centering
    \includegraphics[width=1.0\linewidth]{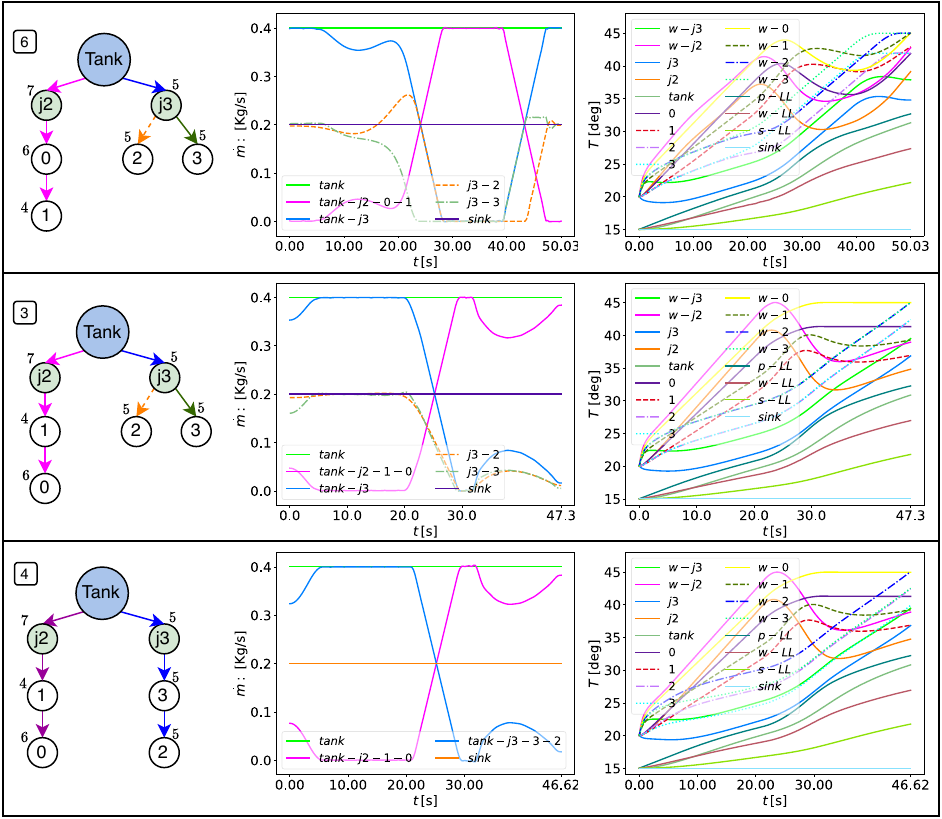}
    \caption{Investigation of the OLOC signals for 3 cases presented in Fig. \ref{fig:investigate_all_configs} and with heat loads = [5,7,6,4,5,5] kW. The multi-split configuration: 6 has the maximum, 4 has the minimum, and 3 has a value in the mid-range of thermal endurance.}
    \label{fig:innerloo_invest_7}
\end{figure}

To understand what happens to each control and state signal when solving the OLOC problem, consider Figs.~\ref{fig:innerloo_invest_5} and \ref{fig:innerloo_invest_7}. In the visualization, the fluid nodes are represented by circles, and the corresponding wall temperatures are indicated in the plot legend using the notation $w$. For instance, the wall node linked to the fluid node $0$ is denoted as $w-0$. All nodes are constrained to a maximum temperature of 45 degrees; if temperature constraints are not active, it is possible to increase thermal endurance. Optimal thermal endurance often occurs when all wall nodes reach the upper-bound at the same time; if one of the nodes reaches the upper-bound sooner than the others, the thermal endurance could be increased (the capacity of the system has not been fully utilized).

Figure~\ref{fig:innerloo_invest_5} compares the optimal control trajectories of the first scenario for three cases: case 0 (best), case 4 (worst), and case 1 (in between). In the best configuration, all wall temperature nodes ($w$) reach the upper-bound at the same time. for configuration 1, only nodes 0, 1, and 3 reach the upper-bound, and for  configuration 4 only nodes 0 and 2 reached to the upper bound. This is reasonable because, while all nodes have the same heat loads, the nodes that are in series get the same flow rate. As a result, the last node will have a higher temperature, as the fluid there has already absorbed the heat from the previous nodes. The results obtained for the flow-rates and all OLOC signals are shown in this figure.

Figure~\ref{fig:innerloo_invest_7} compares the optimal control trajectories for the second scenario where the disturbances are not the same for all nodes. Here, results are shown for case 6 (best), case 4 (worst), and case 3 (in between). In this scenario, the optimal solution is obtained from configuration 6 where junction 2 with the maximum heat-load is in series with nodes 0, 1, see Fig.~\ref{investigate_brute_force_results}(b). In this scenario, configuration 0 is the next optimal solution among the 9 cases. 
Figure~\ref{fig:innerloo_invest_7} also shows the control and state signals. Again for the optimal solution we observe that all the wall nodes reach the upper-bound temperature at the same times so this configuration has the maximum thermal endurance. By comparing configurations 6 and 3 we notice that the only difference between these architectures is the order of the nodes 0 and 1. In configuration 3, in the first branch where all nodes are in series, node 0 is the last node and has a larger disturbances than node 1. When the fluid reaches this node, it is already warmer than the fluid in node 1; therefore, the temperature for that node reaches the upper-bound temperature sooner than node 1. On the other hand, in optimal configuration (6), node 0 is closer to the tank and receives cooler fluid than does node 1. 

In our future work, we plan to expand this study by considering a wider range of graphs and conduct an in-depth analysis. Additionally, we aim to utilize machine learning techniques to extract human interpretable data from the optimization results. This approach will enable us to uncover valuable insights and understand the underlying patterns and relationships in a more systematic and interpretable manner. By leveraging machine learning methods, we can enhance our understanding of the optimization process and gain actionable knowledge that can be applied to further improve the design and performance of such systems.

\subsection{A multi-split configuration with 17 CPHX-nodes}
\label{sec: Multi-split configurations with 17 nodes}
\begin{figure}[ht!]
    \centering
    \includegraphics[width=1.0\linewidth]{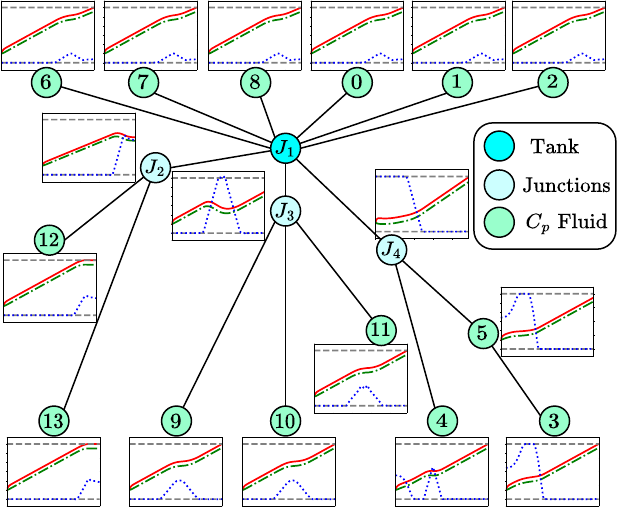}
    \caption{A graph with 18 nodes: 1 tank, 14 CPHXs, and 3 junction CPHXs. Individual plots, adjacent to each node, display the wall temperature (solid red line), fluid temperature (dashed-dot green line), and flow rate (dotted blue line). The plots show temperature (left y-axis) in $^{\circ} \mathrm{C}$ and flow rate in $\mathrm{Kg/s}$ (right y-axis) vs time in $\mathrm{S}$ (x-axis). The dashed gray lines indicates the temperature values of 15 $^{\circ} \mathrm{C}$ and 45 $^{\circ} \mathrm{C}$ , and flow rate of 0 and 0.4 $\mathrm{Kg/s}$. }
    \label{fig:complex_14_nodes}
\end{figure}

In the previous section, smaller-size graphs were studied to facilitate an easier discussion of the results. However, the generated code theoretically has the capability of automatically generating and solving graphs of any size. In this particular case, we considered a graph with 17 heat-exchanger nodes: 14 CPHXs, and 3 junction CPHXs which were added using the spatial locations of the CPHXs. Figure \ref{fig:complex_14_nodes} displays the generated graph. Some CPHXs are directly connected to the tank, while others are connected to the created junctions. All \textit{$C_p$ Fluid} nodes have the same heat load of 4 kW, and the heat loads of junctions $J_2$, $J_3$, and $J_4$ are 3 kW, 4 kW, and 5 kW, respectively. 

Figure \ref{fig:complex_14_nodes} visualises the wall temperature (solid red line), fluid temperature (dashed-dot green line), and flow rate (dotted-blue line) for each node. Additionally, the dashed gray line represents the range of these variables, with temperature ranging from 15 to 45 degrees and flow rate ranging from 0 to 0.4 kg/s. Observing the graph, we can see that the wall temperature of all nodes at the end of each branch (nodes: 0, 1, 2, 3, 4, 6, 7, 8, 9, 10, 11, 12, 13) reaches the upper bound simultaneously. Similarly, the wall temperature of $J_4$, which has the highest heat load among all three junctions, also reaches the upper bound. Moreover, the flow rate of $J_4$ is higher than that of $J_2$ and $J_3$. Additionally, since all nodes have the same heat load, nodes connected to the same junction (or to the tank) exhibit almost identical flow rate signals. As an example, the flow rate values of all nodes within the following three categories are identical: $\{6, 7, 8, 0, 1, 2\}$, $\{9, 10, 11\}$, and $\{12, 13\}$. The objective value of this problem reaches 48.68, indicating the thermal endurance of this system under the given heat load.

\section{Conclusion}
\label{Conclusion}
This article presents the optimal flow control of fluid-based thermal management systems with multi-split configurations. Graph-based modeling is used to generate different configurations and to automatically construct their dynamic equations. In addition, the spatial information of the heat-exchangers is used to define junctions and determine the level of complexity of the systems architecture. The presented generative algorithm can be used for various applications in the domain of configuration design. Next, an open-loop optimization problem is posed to solve the problem, and enumeration is employed to determine the best structure. 

The results include 3 parts. Part 1 compares the results of single and multi-split systems composed of 3 or 4 CPHXs and discusses the results in detail. Part 2 presents and compares the results for multi-split systems composed of 6 CPHXs and gives a comprehensive analysis on the inner-loop optimization. Finally, part 3 studies a larger scale multi-split system made of 17 CPHXs; an analysis of the results obtain is presented. The results show that multi-split configurations result in a better configuration design in some cases. For many optimal cases, all the wall nodes positioned at the end of CPHX system branches reach the upper-bound temperature simultaneously. Additionally, we observed that the optimal configuration depends on the disturbance values. Finally, the results show the possibility of modeling and analysing large system made of various CPHXs.

In the next step, we plan to use a population-based algorithm and compare its results with an enumeration-based framework. Furthermore, our goal is to utilize machine learning techniques to extract a knowledge base interpretable by humans from the optimization data. This knowledge base can then be utilized to guide designers in the process of designing similar systems. Some other future work items include using more sophisticated hydraulic models, performing 3D spatial optimization of the pipe network with simultaneous energy loss minimization, and application of the proposed design framework to some larger industry-relevant applications.

\bibliographystyle{asmejour}   

\bibliography{Main} 



\end{document}